\documentclass[fleqn,usenatbib]{mnras}
\usepackage{newtxtext,newtxmath}
\usepackage[T1]{fontenc}
\usepackage{amsmath}
\usepackage{color}
\DeclareRobustCommand{\VAN}[3]{#2}
\let\VANthebibliography\thebibliography
\def\thebibliography{\DeclareRobustCommand{\VAN}[3]{##3}\VANthebibliography}
\usepackage{graphicx}	
\usepackage{amsmath}	
\usepackage{multirow}
\usepackage{pifont}
\newcommand{\cmark}{\ding{51}}
\newcommand{\xmark}{\ding{55}}

\title[Mass-sheet and similarity transformation degeneracy]{Galaxy lens reconstruction based on strongly lensed gravitational waves: similarity transformation degeneracy and mass-sheet degeneracy}

\author[Jason Poon et al.]{
Jason S.C. Poon$^{1}$\thanks{E-mail: jason.poon@link.cuhk.edu.hk},
Stefano Rinaldi$^{2,3,4,5}$,
Justin Janquart$^{6,7}$,
Harsh Narola$^{6,7}$,
Otto A. Hannuksela$^{1}$
\\
$^{1}$Department of Physics, The Chinese University of Hong Kong, Shatin, New Territories, Hong Kong\\
$^{2}$Dipartimento di Fisica ``E. Fermi'', Università di Pisa, Largo Bruno Pontecorvo 3, 56127 Pisa, Italy\\
$^{3}$INFN, Sezione di Pisa, Largo Bruno Pontecorvo 3, 56127 Pisa, Italy\\
$^{4}$Institut für Theoretische Astrophysik, ZAH, Universität Heidelberg, Albert-Ueberle-Str. 2, 69120 Heidelberg, Germany\\
$^{5}$Dipartimento di Fisica e Astronomia ``G. Galilei'', Università di Padova, Via Marzolo 8, 35121 Padova, Italy\\
$^{6}$Institute for Gravitational and Subatomic Physics (GRASP), Department of Physics, Utrecht University, Princetonplein 1, 3584 CC Utrecht, The Netherlands\\
$^{7}$Nikhef - National Institute for Subatomic Physics, Science Park 105, 1098 XG Amsterdam, The Netherlands
}

\date{Accepted XXX. Received YYY; in original form ZZZ}

\pubyear{2024}

\begin{document}
\label{firstpage}
\pagerange{\pageref{firstpage}--\pageref{lastpage}}
\maketitle

\begin{abstract}
Gravitational wave (GW) galaxy lens reconstruction is a crucial step for many GW lensing science applications. 
However, dark siren GW lensing (\textit{i.e.} lensed GW without observed electromagnetic (EM) counterpart) suffers from similarity transformation degeneracy and mass-sheet degeneracy. 
We review these two degeneracies and discuss their implications on GW-based lens reconstruction and two well-known GW lensing science cases: the Hubble constant measurement and test for modified GW propagation.  
Building upon previous works, our conclusions are:
1) GWs can only infer the scale-free lens mass model parameters, the dimensionless source position, the GW luminosity distance and the time delay scaling (a combination of Einstein radius, lens redshift, and cosmology). 
2) Lens reconstruction (of singular isothermal ellipsoid lens) with only two GW signals is unlikely to yield a complete lens model, while four (three) signals can measure all the above parameters accurately (with large uncertainties). 
3) The similarity transformation degeneracy causes the lens redshift/Einstein radius/cosmology to be degenerate in dark siren measurements. Breaking the degeneracy can be achieved by supplementing the GWs with EM observation of lens redshifts/Einstein radius (source redshift is not required). 
4) The mass-sheet degeneracy causes the GW luminosity distance to be entirely degenerate with a constant mass sheet. 
5) Contrary to expectation, the Hubble constant is degenerate with the mass-sheet even when supplemented with lens reconstruction/redshift/Einstein radius and can only be lifted with lens galaxy velocity dispersion measurement, while modified GW propagation test discussed in prior literature is unaffected by the degeneracy. 
These properties highlight the need for GW observations to be supplemented by EM observations, which could become accessible through a lens archival search or a rapid EM follow-up. 
\end{abstract}


\section{Introduction}
The gravitational lensing of light has been captured by electromagnetic (EM) telescopes and studied for many decades, and it has become a standard yet powerful tool in modern astronomy \citep{1998LRR.....1...12W, Bartelmann_2010}. Its mature development and methodology allow scientists to probe the universe, with some of its science applications including cosmological studies (such as Hubble constant measurement \citep{williams1997measurement}), dark matter studies (for instance, testing if massive compact halo object (MACHO) is a possible candidate \citep{Basak_2022}), lens galaxy density profile and substructures studies, as well as galaxy cluster structure studies \citep{Bartelmann_2010}. 

Meanwhile, according to general relativity, gravitational waves (GWs) are expected to be lensed like light when they propagate near massive objects \citep{2002A&A...394..749D}. GW science itself is undergoing rapid development since its first direct detection by Laser Interferometer Gravitational-Wave Observatory (LIGO) in 2015 \citep{PhysRevLett.116.061102}: it opened up new windows to probe and study countless scientific problems, including neutron star and black hole physics (such as their structure and formation channels), testing general relativity and cosmological studies such as expansion of universe and dark energy \citep{Sathyaprakash_2009, Lattimer_2012, Gair_2013, Yunes_2013, Will_2014, Berti_2015, lasky_2015, Abbott_2019a, Abbott_2019, Baiotti_2019, Abbott_2021, 2021NatRP...3..344B, Abbott_2022}. Thanks to technological advancement, current GW detectors (\textit{i.e.} the second generation detectors), such as aLIGO \citep{2015}, aVIRGO \citep{Acernese_2014} and KAGRA \citep{10.1093/ptep/ptaa125, PhysRevD.88.043007, Somiya_2012}, are undergoing substantial upgrades \citep{galaxies10010036, bagnasco2023ligovirgokagra}, and more new and advanced detectors are being planned, such as the ground-based detector Einstein Telescope \citep{2010CQGra..27s4002P} and Cosmic Explorer \citep{reitze2019cosmic}, as well as the space-based detector Laser Interferometer Space Antenna (LISA) \citep{amaroseoane2017laser}. 

One avenue that has garnered recent interest is the gravitational lensing of GWs. Several groups have forecasted that, at their design sensitivity, second-generation detectors would detect strongly-lensed GW with a detection rate of around one such event per year, assuming standard formation channels \citep{Ng_2018,Li_2018, Ng_2018, Oguri_2018, Wierda_2021, Xu_2022,smith2023discovering,gupta2023characterizing}. These forecasts have helped motivate a recent effort to search for GW lensing (see~\cite {Hannuksela_2019, McIsaac_2020, dai2020search, Liu_2021, PhysRevD.104.103529, 2021ApJ...923...14A, 2023arXiv230408393T, Janquart_2023}) and develop scientific applications for them. 

Indeed, if detected, GW lensing might allow several applications in fundamental physics, astrophysics, and cosmology. For example, GW lensing might allow sub-arcsecond localization of GW sources \citep{Hannuksela_2020, wempe2022lensing}. Other applications include tests of modified GW propagation \citep{Finke_2021, narola2023modified}, testing the speed of gravity \citep{Collett_2017, Fan_2017}, GW polarization test \citep{Smith_2017, Goyal_2021, hernandez2022measuring}, other GW properties \citep{Yang_2019, Mukherjee_2020, Mukherjee_2020a, Chung_2021, Goyal_2021, hernandez2022measuring, iacovelli2022modified, colaço2023forecasts}, cosmography \citep{Sereno_2010, Liao_2017, Li_2019, Yang_2019, Hou_2021, Cao_2022, Wang_2022, Wu_2023} including high-redshift cosmography \citep{Jana_2023}, probing dark matter, subhalos, and compact objects \citep{Takahashi_2006, Dai_2018, Liao_2018, Diego_2020, Oguri_2020, Choi_2021, Wang_2021, Basak_2022, Cao_2022, Guo_2022, Oguri_2022, Fairbairn_2023, liu2023gravitational, Liu_2023, seo2023inferring, tambalo2023gravitational}, associating fast radio burst source and GW source \citep{singh2023dejavu}, constraining binary system formation channel \citep{Mukherjee_2021}. Indeed, detections could enable a plethora of applications.

For many science applications, such as those in cosmography and tests of modified GW propagation, "reconstruction" of the gravitational lens is one of the crucial steps in the analysis. In particular, lensed GWs encode information about the mass profile of the objects that gravitationally lensed them. The attempt to reconstruct this lens mass profile and other extrinsic properties of the lens is collectively called "lens reconstruction." Such reconstructions allow one to study the lens model itself, including its profile and the possible substructures within. 

However, strong GW lensing suffers from two major degeneracies which limit these reconstructions. The first one is the similarity transformation degeneracy\citep{1988ApJ...327..693G}, which refers to the fact that lensed GWs cannot disentangle the size of the gravitational lens from the angular diameter distances. \textit{I.e.} the lens size is degenerate with the distances between the lens, source and observer. The second degeneracy is mass-sheet degeneracy \citep{1985ApJ...289L...1F}, which refers to the fact that adding a "constant mass sheet" to the lens profile will leave the image positions and flux ratios invariant. In GW observations, as we will see, such a constant mass sheet is degenerate with the inferred luminosity distance and the overall time delay. These degeneracies limit the scientific applications of GW lensing and what can be done with GWs alone.\footnote{Note that wave optics lensing and GWs supplemented by EM information are an exception \citep{Hannuksela_2020,Cremonese_2021} (Chen et al., in prep.).}

The similarity transformation degeneracy was originally reviewed by \citet{1988ApJ...327..693G} and \citet{Saha_2000}, but the effect has yet to be discussed significantly in the context of lensed GW observations. In this work, we would introduce this degeneracy, but more in the context of GWs, based on their findings. Due to the similarity transformation degeneracy, lens systems with different redshifts, sizes and cosmological parameters (such as the Hubble constant) can produce identical lensing observables. In particular, the size of the lens is degenerate with the distances in the source-lens-observer system. The degeneracy can be broken when the "size" of the lens can be obtained from complementary EM observation. For example, a complementary telescope image of the lensing object can break the degeneracy (because the lens size and a redshift of the source and the lens can be directly observed in the telescope). Such a telescope image might be available if the lens system is located by an archival strong lens catalog\footnote{Such as, with \texttt{lenscat} https://github.com/lenscat/lenscat.} search \citep{Hannuksela_2020,wempe2022lensing,shan2023microlensing}, which could be performed without an EM counterpart. Alternatively, a rapid EM followup, if an EM counterpart to the event is available, would also allow us to retrieve such a telescope image \citep{smith2023discovering}. If no supplementary EM information is available, the degeneracy limits GW applications. Thus, it is crucial to understand the effect. 

The second degeneracy is the mass-sheet degeneracy, which leaves the image positions, flux ratios, and time-delay ratios unchanged when a uniform and infinitely-spanning mass sheet is added to the lens model \citep{1985ApJ...289L...1F}. Such mass sheet can approximate scenarios where a galaxy lens is embedded in low-mass galaxy cluster, there are massive structures along the line of sight, or the lens has a positive power-law slope that differs from an isothermal profile~\citep{1985ApJ...289L...1F, Schneider_2013, Birrer_2016,oguri2019strong}. It has been a known degeneracy affecting strong lens modelling (of light) for many decades \citep{Schneider_2013}, and it affects GW lens reconstruction the same way. In particular, the constant mass sheet will be degenerate with the luminosity distance and time delay measured by the GWs. However, as we will see, complementary information about the source redshift can alleviate the degeneracy, but not in cosmological applications. Even when the source redshift is known, the degeneracy causes the Hubble constant to be degenerate with the mass-sheet density.  Another intriguing way out of the degeneracy is by measuring the wave optics effects when the lens size is comparable to the wavelength of the GW \citep{Cremonese_2021} (Chen et al., in prep.). Without wave optics effects or complementary EM redshift information, the degeneracy limits many GW applications, so it is important to understand the effect. 

This work reviews the idea of GW-based lens reconstruction/modelling of galaxy-type lenses. We also discuss the nature and effect of the two degeneracies and how one should consider them in the GW lensing context. In Section \ref{S2}, we first review the basics of strong gravitational lensing for both light and GWs, before focusing more on its effect on GWs. Section \ref{S3} explains lens reconstruction for lensed GW signal using a Bayesian framework. Section \ref{S4} introduces the similarity transformation degeneracy and discusses its impact on GW-based lens reconstruction. In Section \ref{S5}, we discuss the mass-sheet degeneracy problem in the context of GW lensing. In Section \ref{S6}, we present the impact of the degeneracies on the test for modified GW propagation using lens reconstruction. We also discuss their implications for various other science cases.

Throughout the paper, we assume a flat $\Lambda$CDM universe with cosmology from Plank18 data: $\Omega_m=0.308, \Omega_K=0, \Omega_\Lambda=0.692$ \citep{2020}, with $H_0$ being a free (and unknown) parameter unless stated.

\section{Basics of Gravitational Wave Strong Lensing} \label{S2}
In section \ref{S2.1}, we will first review the basics of strong gravitational lensing in geometrical optics, which applies to both EM waves and GWs as long as the wavelength is much shorter than the scale of the lens object. If the wavelength is comparable or longer than the characteristic lens scale, wave optics (which includes diffraction and interference effect) must be considered \citep{10.1143/PTPS.133.137, Takahashi_2003}. However, because the typical size of strong galaxy lenses is much larger than the wavelength of GWs measurable with ground-based detectors, we limit our discussion to the geometrical optics effects. We will also neglect microlensing effects.\footnote{Note that microlensing \emph{could} become relevant in actual galaxy lensing scenarios and will likely require further investigation \citep{diego2019observational,cheung2021stellar,mishra2021gravitational,meena2022gravitational,yeung2023detectability,mishra2024exploring}.} Then, in section \ref{S2.2}, we will review the effect of strong gravitational lensing on GWs, especially on the observation end.

\subsection{Strong gravitational lensing} \label{S2.1}
\begin{figure}
\centering
\includegraphics[width=\columnwidth]{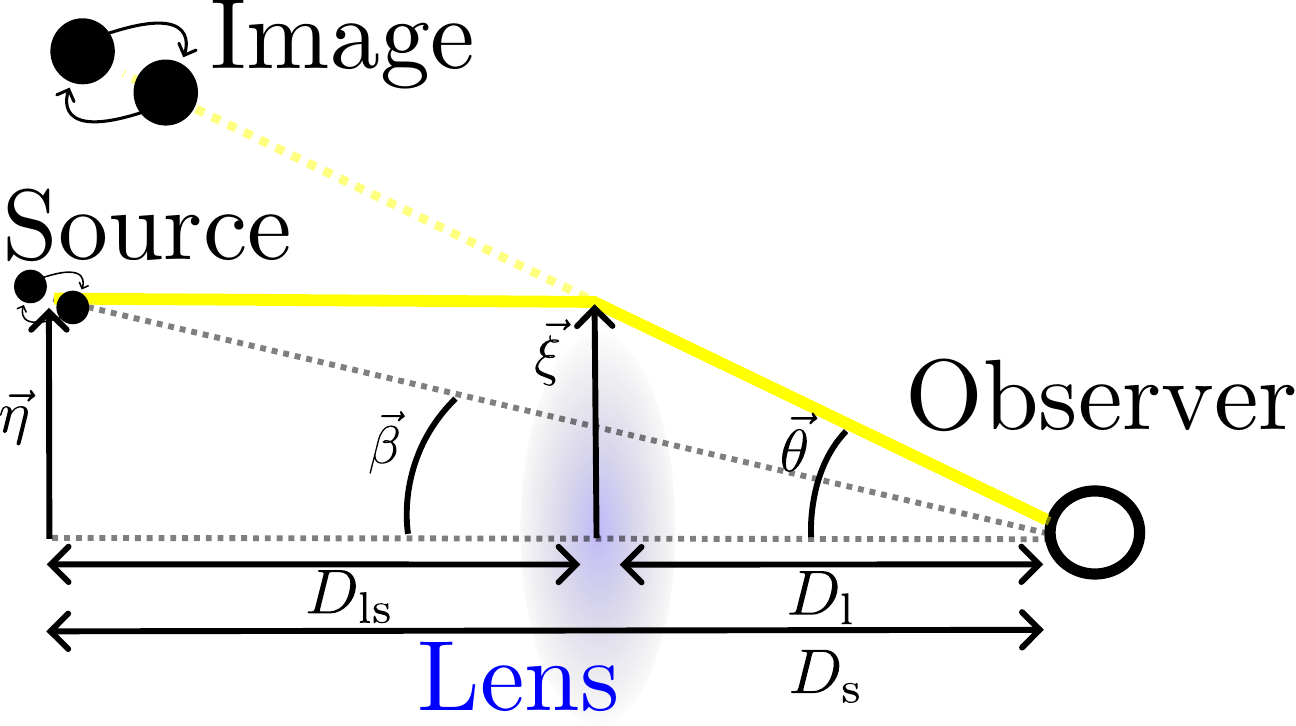}
\caption{Geometry of a lens system: $D_l, D_s, D_{ls}$ are the angular diameter distances from observer to lens, observer to source and lens to source, respectively. Aligning our principal axis with the centre of the lens, $\vec{\beta}, \vec{\theta}$ are the angular position of the source and image respectively, with $\vec{\eta}=D_s \vec{\beta}$ and $\vec{\xi} = D_l \vec{\theta}$ being the (Cartesian) source and image position(s) (on source plane and lens plane) respectively.}
\label{fig:lens_geometry}
\centering
\end{figure}

Fig. \ref{fig:lens_geometry} illustrates the typical setup of a lens system. We denote the angular diameter distance from the observer to the lens, observer to source and lens to source as $D_l, D_s, D_{ls}$ respectively. The redshift of lens and source are denoted as $z_l$ and $z_s$, respectively. We also denote the angular position of the source and image by $\vec{\beta}$ and $ \vec{\theta}$, respectively. The characteristic scale of the lens is the angular Einstein radius $\theta_E$ (some literature may refer to this as the Einstein angle), which depends on the lens mass and the angular diameter distances in all generality.

In this work, we should work in dimensionless coordinates of the lens system: 
we define the dimensionless source and image positions as $\vec{y}=(y_1, y_2)=\vec{\beta}/\theta_E$ and $\vec{x}=(x_1,x_2)=\vec{\theta}/\theta_E$ respectively. 
The dimensionless deflection/lensing potential $\Psi$ (deflection potential scaled by $\theta_E^2$) is then defined by
\begin{equation}
    \Psi(\vec{x})=\frac{1}{\pi}\int{\kappa(\vec{x}') \text{ln}|\vec{x}-\vec{x}'| d^2 x'},
\end{equation}
where the convergence $\kappa$ is defined by
\begin{equation}
    \kappa(\vec{x})=\frac{\Sigma(\vec{\xi}=D_l \theta_E \vec{x})}{\Sigma_{cr}},
\end{equation}
where $\Sigma(\vec{\xi})$ is the surface mass density of the lens (in cartesian coordinate $\vec{\xi}$ on the lens plane) and $\Sigma_{cr}={c^2 D_s}/{4 \pi G D_l D_{ls}}$ is the critical density \citep{Takahashi_2003, tambalo2023gravitational}. 
This particular definition highlights the fact that $\Psi(\vec{x})$ does not contain scaling information, such as the total mass of the lens. 
As we will see, the notation is advantageous because the dimensionless quantities are unaffected by the similarity transformation degeneracy.

The lens equation in dimensionless form can then be written as
\begin{equation}
    \vec{y} = \vec{x} - \nabla_{\vec{x}} \Psi(\vec{x}),
    \label{dimensionless lens eq}
\end{equation}
which solving for $\vec{x}$ will give us the (dimensionless) image position(s) under a known source position and lens model (there can be multiple images in general).

Meanwhile, the time delay $t_d$ of an image compared to an unlensed ray is given by
\begin{equation}
    t_d(\vec{x}) = \frac{1+z_l}{c}\frac{D_l D_s}{D_{ls}} \theta_E^2\left[\frac{|\vec{x}-\vec{y}|^2}{2}-\Psi(\vec{x})\right],
    \label{td eq}
\end{equation}
where ${|\vec{x}-\vec{y}|^2}/{2}$ accounts for the delay due to extra distance travelled by the lensed ray, and $\Psi(\vec{x})$ accounts for the delay due to gravitational time dilation effect \citep{Takahashi_2003, Bulashenko_2022}. Also, the magnification $\mu$ of the image is given by
\begin{equation}
    \mu(\vec{x})=\left[\det \left(\frac{\partial \vec{y}}{\partial{\vec{x}}}\right)\right]^{-1},
    \label{magnification eq}
\end{equation}
where $\frac{\partial \vec{y}}{\partial{\vec{x}}}$ is given by

\begin{equation}
    \frac{\partial \vec{y}}{\partial{\vec{x}}}=
\begin{pmatrix}
1-\frac{\partial^2 \Psi}{\partial x_1^2} & -\frac{\partial^2 \Psi}{\partial x_1 \partial x_2}\\
-\frac{\partial^2 \Psi}{\partial x_2 \partial x_1} & 1-\frac{\partial^2 \Psi}{\partial x_2^2}
\end{pmatrix}.
\end{equation}
We note that the magnification of an image is scale-free, which means it only depends on the dimensionless system.

\subsection{Effect of strong lensing on gravitational waves} \label{S2.2}
Denoting $\tilde{h}_0(f)$ the original unlensed GW waveform in the frequency domain and $\tilde{h}_L(f)$ the observed lensed ones, the relation between the two waveforms is
\begin{equation}
    \tilde{h}_L(f) = \tilde{h}_0(f) \sum_j |\mu(\vec{x}_j)|^{1/2} \exp{\left[2\pi i f t_d(\vec{x}_j) - i \pi n_j \text{sign}(f)/2\right]},
\end{equation}
where the index $j$ denotes the $j$-th lensed image/signal, and $n_j$ is its Morse index, with $n_j=\{0, 1, 2\}$ (referred as type I, II and III image) when the image $\vec{x}_j$ is at a local minimum, saddle point and local maximum of the time delay surface $t_d(\vec{x}_j)$ respectively \citep{Takahashi_2003}. We also note that the emergence of Morse indices results from stationary phase approximation in geometrical optics \citep{10.1143/PTPS.133.137}.

Therefore, in time domain, if $n_j=0$ or $2$, the lensed signal $h_L(t)$ and unlensed signal $h_0(t)$ are related by
\begin{equation}
    h_L(t) = \sum_j |\mu(\vec{x}_j)|^{1/2} h_0\left[t-t_d(\vec{x}_j)\right](-1)^{n_j/2},
\end{equation}
and if $n_j=1$, the relation is
\begin{equation}
    h_L(t) = \sum_j |\mu(\vec{x}_j)|^{1/2}\int^{+\infty}_{-\infty}{-\tilde{h}_0(f)\text{sign}(f) i e^{-i2\pi f\left[t-t_d(\vec{x}_j)\right]}},
\end{equation}
which different images arrive at different times and have different amplitudes (due to magnification) \citep{dai2017waveforms,Ezquiaga_2021}.\footnote{If the GW system is symmetric and contains no higher-order modes, the Morse phase is absorbed into the phase of coalescence \citep{dai2017waveforms,Ezquiaga_2021}.} 
However, they will have the same frequency-domain evolution (up to magnification and the phase shift (Morse phase) caused by the Morse index) \citep{Takahashi_2003}, which is what strong lensing GW searches rely on, trying to find events with compatible characteristics \citep{haris2018identifying, Goyal_2021a, Janquart_2021, Kim_2021, janquart2022golum, More_2022, Janquart_2023a, li2023teslax, Li_2023, chakraborty2024glance}. Such searches have already been carried out on real data, but there is no universally accepted evidence for GW lensing detections yet, although many authors have considered the situation \citep{Hannuksela_2019, Liu_2021,2021ApJ...923...14A, Bianconi_2023, PhysRevD.104.103529, Janquart_2023, 2023arXiv230408393T}.

For a lens and GW source system, assuming we successfully identified all of its lensed images (\textit{i.e.} we identify the lensed GW signals from the detector data stream), we can then measure the lensing observables, namely the magnifications, time delays, and Morse phases of the strong lensing images. Firstly, since GW is a standard siren \citep{Holz_2005}, from the lensed GW images (denote the i-th image by subscript hereafter), we can observe the effective (or apparent) luminosity distance $d_{L,i}^\text{eff}$, which is related to the image magnification $\mu_i$ by
\begin{equation}
    d_{L,i}^\text{eff}=\frac{d_L}{\sqrt{|\mu_i|}},
    \label{d_l^eff eq}
\end{equation}
where $d_L$ is the true luminosity distance without the lensing effect. Note that, in many traditional strong lensing scenarios, the absolute flux of the light source is not know.\footnote{Exceptions to this would be strong lensing of standard candles such as supernovae.} In such scenarios, one can only measure the relative magnification ratios ${\mu_i}/{\mu_j}$ between two images (from the flux ratios) \citep{alma991039610282103407}. The absolute flux or the effective luminosity distance are not accessible. However, in lensed GWs (and standard candles), one can measure the effective luminosity distance. This allows for an extra degree of information on $d_L$.
Secondly, in GW lensing, the relative time delay between two images, $\Delta t_{d,ij} = t_{d,i} - t_{d,j}$ can also be measured with submillisecond accuracy \citep{Fan_2017}. Hence, the precise time delay can further constrain the lens model.
Thirdly, by comparing the waveform of the lensed GW signals, we can also observe the relative Morse index (or difference in Morse index) $\Delta n_{ij} = n_i - n_j$ between different images \citep{Janquart_2021b}.
The direct GW observables are then the effective luminosity distances, relative time delays, and relative (or in some cases,  absolute \footnote{Often we can only accurately resolve the \emph{relative} Morse phase due to a degeneracy between the Morse phase and GW coalescence phase \citep{dai2020search,Lo_2023,Janquart_2021b}, which however can be lifted in the presence of higher-order modes.}) Morse phases of the strong lensing images. 

\begin{figure*}
\centering
\includegraphics[scale=0.47]{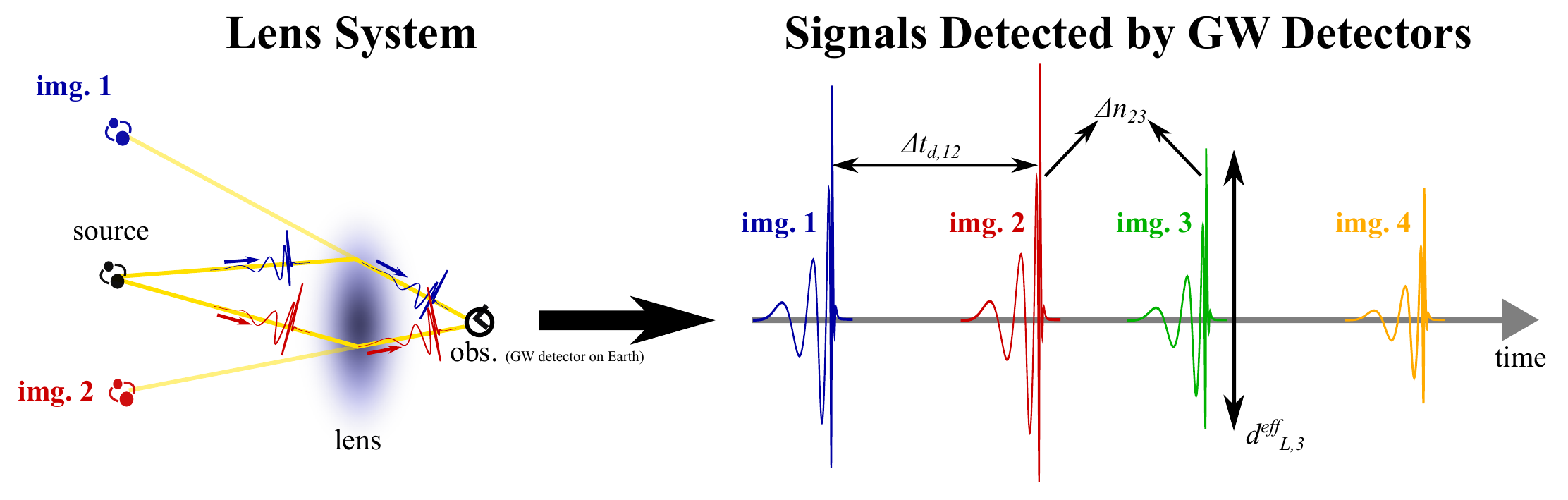}
\caption{Illustration of dark siren GW lensing: the left panel shows the setup of the lens system, which the GWs will arrive the observer (obs.) at different time, with different amplitude and different Morse index. The observables in a dark siren GW lensing scenario are shown on the right panel: from the data recorded by the GW detector, one can measure the relative time delay $\Delta t_{d,ij}$ and the relative Morse index $\Delta n_{ij}$ between images, and since GW is a standard siren, from the amplitude of the GW signal (together with waveform analysis), one can measure the effective/apparent luminosity distance $d_{L,i}^\text{eff}$ of the images.}
\label{fig:dark_siren_GW_lensing_illustration}
\centering
\end{figure*}

In many traditional strong lensing scenarios, the telescope angular resolution is sufficiently good to locate and resolve the multiple strong lensing images. However, due to the broad sky localization area of a GW signal (order of ten to hundreds of square degrees in the near future \citep{Pankow_2018}), unlike in many traditional lensing observation, we would not be able to directly localize the precise sky location of the GW signals/images individually from the GW data alone \citep{Hannuksela_2020}. Therefore, we will henceforth name ``dark siren GW lensing'' as the scenario in which we only have observation of lensed GW from the GW detectors but do not have any EM information on the lens and source system due to lack of EM counterpart for the compact binary merger or broad sky localization \footnote{However, we note that it is still possible to identify the dark siren GW lens system to the same lens system that lensed the GW source's host galaxy if it is observed by EM telescopes. The identification would require statistical comparison of the lensing observables measured in the two channels. It will be discussed in Section \ref{S3}.}. All the GW data will be denoted $\vec{d}_{GW}$. The data includes information regarding the images' effective luminosity distance $d_{L,i}^\text{eff}$, the relative time delay between image pairs $\Delta t_{d,ij}$ and the relative Morse index $\Delta n_{ij}$ between different images (Fig. \ref{fig:dark_siren_GW_lensing_illustration}).

\section{Lens Reconstruction Using Bayesian Analysis} \label{S3}
This section reviews the basics of lens reconstruction using a Bayesian analysis framework for GW lensing. In particular, we will focus on galaxy lenses. We note that similar methods have been adopted by various works previously, for example, \citet{Hannuksela_2020, wempe2022lensing, Wright_2022, seo2023inferring, wright2023determination}.

There are various ways to model lenses, with some approaches being parametric and others non-parametric. Based on strong galaxy lensing quasar measurements, it was realized early on that even relatively simple smooth lens profiles (such as the singular isothermal ellipsoid (SIE) model \citep{1994A&A...284..285K}) fitted the time delays of quasars relatively well, but the fluxes were difficult to be reproduced due to so-called flux-ratio anomalies \citep{1998MNRAS.295..587M, Metcalf_2001}. Since these early works, more sophisticated parametric models have been developed to study strong galaxy lenses, accounting for additional degrees of freedom such as parameters to describe substructures, shearing, convergence, and non-singular density profiles \citep{alma991039610282103407}. However, to first order (and neglecting shears), one could assume that the general morphology of many strong galaxy lenses is modelled reasonably well using a singular isothermal ellipsoid (SIE) when supplemented with conservative modelling errors to account for the additional degrees of freedom not included in the SIE model \citep{Keeton_1997}.\footnote{Ultimately, corrections to the SIE model are necessary; for example, the addition of shear will introduce a larger spread on the expected, allowed magnification values, which would need to be marginalised over. However, to understand the degeneracies and the general idea behind strong GW lens modelling, an SIE model with generous modelling uncertainties is expected to provide a fair qualitative description and starting point.}

After choosing a lens model, the gravitational lens is characterized by the parameters of the lens. We denote the lens parameters $\vec{\Phi}_l$. The lens parameters encode all information strictly related to the lens, including the lens redshift $z_l$ and the parameters describing the lens surface mass density $\Sigma$ (\textit{i.e.} the mass distribution of the lens).  For instance, for point mass lens, the main lens parameter would be its mass $M_\text{PML}$ (\textit{i.e.} $\vec{\Phi}_l=(z_l,M_\text{PML})$). For SIE, we have $\vec{\Phi}_l=(z_l,f,v)$ where $f$ is the lens' minor to major axis ratio and $v$ is the velocity dispersion. In this work, we restrict our discussion to the SIE model as an approximate representation of a galaxy lens.\footnote{The effect of making the lens mass profile more complex will be discussed in a later section.} 

The second set of parameters encompasses the GW source information $\vec{\Phi}_s$, including the source position $\vec{\beta}$ and the source redshift $z_s$. Since we only focus on the effects induced by lensing but not the nature of the GW waveform, we do not need the binary orbital information (such as its chirp mass and mass ratio).\footnote{The other GW waveform parameters are nevertheless important, but they do not enter into lens modelling unless there is a relationship between, for example, the total mass of the binary and the brightness of the galaxy (see \citet{wempe2022lensing}, for more information).} The last set of parameters contains the cosmological information $\vec{\Phi}_{c}$. For simplicity, we only include the Hubble constant $H_0$ as a free parameter here (and we also only consider a flat $\Lambda$CDM universe).\footnote{In general, one could also include radiation energy density $\Omega_r$, matter density $\Omega_m$ and dark energy density $\Omega_\Lambda$ as free parameters. However, we limit  the analysis to only $H_0$ for simplicity.} Table \ref{table: lens system parameters} summarises the different parameter sets.

\begin{table}
\centering
\begin{tabular}{|c | c | c | c|} 
 \hline
 Type: & $\vec{\Phi}_l$ (lens) & $\vec{\Phi}_s$ (source) & $\vec{\Phi}_{c}$ (cosmology) \\ [0.5ex] 
 \hline
 Free parameters: & $z_l, \Sigma$ & $z_s, \vec{\beta}$ & $H_0$ \\ 
 \hline
\end{tabular}
\caption{Three types of parameters in a GW lens system: the lens parameters $\vec{\Phi}_l$, the GW source parameter $\vec{\Phi}_s$ and the cosmological parameters $\vec{\Phi}_{c}$. Within $\vec{\Phi}_l$, $z_l$ is the lens redshift and $\Sigma$ is the mass distribution of the lens, for which one should assume a mass model (in this work, we assume the SIE model with conservative time-delay and magnification modelling uncertainties, which characterize the mass distribution by two parameters: the minor to major axis ratio $f$ and galaxy velocity dispersion $v$). For $\vec{\Phi}_s$, note that we are only considering the effect induced by lensing (which are magnification and delayed arrival time of the signal, but not the detail of the signal or waveform itself), hence for the GW source, only its redshift $z_s$ and source position $\vec{\beta}$ are relevant. For $\vec{\Phi}_{c}$, in this work, we only consider the Hubble constant $H_0$ as a free parameter in our flat $\Lambda$CDM universe.
}
\label{table: lens system parameters}
\end{table}

The idea of lens reconstruction is to reconstruct the lens information $\vec{\Phi}_l$ based on given observables: in the case of dark siren GW lensing, the observables are the effective luminosity distances $d_{L,i}^\text{eff}$, relative time delays $\Delta t_{d,ij}$ and relative Morse index (Morse index difference) $\Delta n_{ij}$ (Fig. \ref{fig:dark_siren_GW_lensing_illustration}). If the lens system is located through, e.g., archival search, then a telescope image of the source galaxy would be available. The telescope image would provide the flux ratios (and therefore the magnification ratios ${\mu_i}/{\mu_j}$) as a function of the image positions $\vec{\theta}_i$, $\vec{\theta}_j$. In the case that an EM counterpart is available, the GW image positions and fluxes could also be retrieved. Other supplementary information would include the photometric/spectroscopic redshifts of the lens galaxy $z_l$ and the source galaxy $z_s$, and the measurement of the galaxy lens velocity dispersion $\sigma_v$. Nevertheless, without supplementary EM information, the lens reconstruction can only rely on the effective luminosity distances, time delays, and relative Morse phases retrieved from the GWs. 

Aside from the lens profile, inferring source and cosmological information $\vec{\Phi}_s \text{ and }  \vec{\Phi}_{c}$ is also interesting. Lens reconstruction turns out to be a very important tool or middle-step in various science applications of lensing, including sub-arcsecond localization of GW source, measuring Hubble constant and testing modified gravity effect on the propagation of GWs \citep{Hannuksela_2020, Finke_2021, narola2023modified}. However, as we will see, even in the presence of a supplementary, high-resolution telescope image and photometric/spectroscopic redshifts, the mass-sheet degeneracy will limit cosmological applications, while, contrary to expectations, the degeneracy does not affect modified GW propagation tests. 

We use standard Bayesian parameter estimation to reconstruct the lens. In particular, our goal is to infer the posterior distribution of the lens and source parameters, and also cosmological information given measured data: $$P(\vec{\Phi}_l, \vec{\Phi}_s, \vec{\Phi}_{c}|\vec{d}_{GW}),$$ where $\vec{d}_{GW} = \{d_{L,i}^\text{eff}, \Delta t_{d,ij}, \Delta n_{ij}\}$ is the GW lensing observables that we measured. For simplicity, denote all system parameters as $\vec{\Phi}_{all}=\{\vec{\Phi}_l, \vec{\Phi}_s, \vec{\Phi}_{c}\}$. As mentioned, in this work, we will adopt parametric models\footnote{Primarily the SIE lens model.} in reconstructing the lens. Hence, the dimensionless deflection potential $\Psi(\vec{x})$ and Einstein radius $\theta_E$ will be parametric\footnote{\textit{I.e.} $\Psi(\vec{x})\rightarrow \Psi(\vec{x};\vec{\Phi}_l)$ and $\theta_E\rightarrow \theta_E(\sigma_v, D_l, D_{ls}, D_s)$.}. So, we denote the parameters that characterise the selected dimensionless lens profile as $\vec{\Phi}_\Psi$ (for example, for an SIE lens, $\vec{\Phi}_\Psi=f$), and now the lens parameters are represented by $\vec{\Phi}_l=(\theta_E,\vec{\Phi}_\Psi, z_l)$. Then, the system parameters are effectively $\vec{\Phi}_{all}=(\vec{\beta}, z_s, \theta_E,\vec{\Phi}_\Psi, z_l, H_0)$. As a word of caution, however, some of these parameters will turn out to be degenerate with one another when only GW observations (\textit{i.e.} dark siren) are available. We will nevertheless first proceed with the Bayesian formulation using the full set of parameters. 

The Bayes theorem allows one to formulate the inference problem. In particular, using the Bayes theorem, we can write
\begin{equation}
    P(\vec{\Phi}_{all}|\vec{d}_{GW}) = \frac{P(\vec{d}_{GW}|\vec{\Phi}_{all})P(\vec{\Phi}_{all})}{P(\vec{d}_{GW})},
\end{equation}
where $P(\vec{d}_{GW}|\vec{\Phi}_{all})$ is the likelihood of obtaining measured data given system parameters, $P(\vec{\Phi}_{all})$ is the prior (probability distribution $\vec{\Phi}_{all}$ set by prior knowledge) and $P(\vec{d}_{GW})$ is the evidence. One can obtain the posterior $P(\vec{\Phi}_{all}|\vec{d}_{GW})$ by nested sampling.\footnote{Nested sampling can be be used to infer the posterior parameters as long as a likelihood $P(\vec{d}_{GW}|\vec{\Phi}_{all})$ (which included the measured data $\vec{d}_{GW}$), and a prior $P(\vec{\Phi}_{all})$ is defined.} The nested sampling procedure would then output both the posterior $P(\vec{\Phi}_{all}|\vec{d}_{GW})$ and the evidence $P(\vec{d}_{GW})$. The evidence is important for model comparison (say, comparing the point mass lens model against the SIS lens model as demonstrated in \citet{wright2023determination}, or comparing the SIE lens model versus PEMD (power law elliptical mass density profile) lens model), which we do not do here. We focus on the lens reconstruction (inference of the lensing parameters) instead of model selection/comparison.

We note that the likelihood $P(\vec{d}_{GW}|\vec{\Phi}_{all})$ cannot be written down directly, as the GW data only contains the measured image properties $d_{L,i}^\text{eff}, \Delta t_{d,ij}$ and $\Delta n_{ij}$, but no information about $\vec{\Phi}_{all}$ directly. Hence, we would need a mapping to relate $\vec{d}_{GW}$ and $\vec{\Phi}_{all}$. We denote the image properties as $\vec{\nu}_{im}=(d_{L,i}^\text{eff}, \Delta t_{d,ij}, \Delta n_{ij})$. Note that image properties are functions of the system parameters: $\vec{\nu}_{im}=\vec{\nu}_{im}(\vec{\Phi}_{all})$. Assuming a suitable noise model, we can obtain $P(\vec{d}_{GW}|\vec{\nu}_{im})$ easily (as demonstrated in previous works such as \citet{Liu_2021, Janquart_2023, Lo_2023}). To obtain the likelihood $P(\vec{d}_{GW}|\vec{\Phi}_{all})$, we note that $P(\vec{d}_{GW}|\vec{\Phi}_{all})=\int{P(\vec{d}_{GW}|\vec{\nu}_{im})P(\vec{\nu}_{im}|\vec{\Phi}_{all}) d\vec{\nu}_{im}}$. Since the mapping from $\vec{\Phi}_{all}$ to $\vec{\nu}_{im}$ is a one to one mapping, we have $P(\vec{\nu}_{im}|\vec{\Phi}_{all}) = \delta(\vec{\nu}_{im}-\vec{\nu}_{im}(\vec{\Phi}_{all}))$. Therefore, to sample the likelihood, one can set all the system parameters $\vec{\Phi}_{all}$ as free parameters, then solve the image positions from the lens equation (Eq.(\ref{dimensionless lens eq})), then calculate the image properties $\vec{\nu}_{im}(\vec{\Phi}_{all})$ using Eq.(\ref{td eq}), Eq.(\ref{magnification eq}) and Eq.(\ref{d_l^eff eq}), and relate it to the measured data by the likelihood $P(\vec{d}_{GW}|\vec{\nu}_{im})$.

As mentioned, in dark siren GW lensing observation, due to various degeneracies, we would not be able to infer the posterior $P(\vec{\Phi}_{all}|\vec{d}_{GW})$ of all the parameters $\vec{\Phi}_{all}$ in the lens system. Instead, only posterior distributions of some of all free parameters and a combination of some free parameters can be inferred. Nevertheless, the idea of lens reconstruction in the Bayesian framework discussed here remains unchanged.

After dark siren GW lens reconstruction, one could attempt to localize the corresponding lens-source system in a follow-up archival search \citep{Hannuksela_2020}. Such a joint analysis could lead to a positive identification of the lens system in EM channels if the lens galaxy is found and it is sufficiently unique such that a joint analysis between the EM lens reconstruction and the GW data could rule out a sufficient number of lenses. The methodology was demonstrated in \citet{Hannuksela_2020} and \citet{wempe2022lensing}. If the source and host galaxy are localized, one could obtain the redshifts of the source and the lens through photometric/spectroscopic measurement, and a supplementary lens reconstruction that could resolve some of the quantities that cannot be resolved from GWs, such as the Einstein radius, in addition to giving a supplementary constraints on the other lensing parameters (such as the axis ratio). However, the localization starts with a dark siren GW lens reconstruction analysis. Furthermore, understanding what type of information GW-based strong lensing reconstructions can provide by themselves and what type of caveats or degeneracies they are subject to without supplementary information, will help in outlining complementary avenues between the two. Thus, in this work, we perform lens reconstruction based on (injected) dark siren lensed GW observations. 

To infer the lens parameters, we use the GW Bayesian inference tool \texttt{Bilby} \citep{Ashton_2019} and the nested sampler \texttt{PyMultiNest} \citep{Buchner_2014}. To model the gravitational lensing effects, we use the multi-purpose gravitational lensing tool \texttt{lenstronomy} \citep{Birrer_2018}.

\section{Similarity Transformation Degeneracy} \label{S4}

The similarity transformation degeneracy is a degeneracy between the (angular diameter) distances in the lens system and the angular scale (\textit{i.e.} the Einstein radius $\theta_E$) of the lens, and the main reason for this degeneracy is due to the scale-free nature of many lensing observables (such as the magnification and time delay ratio) \citep{Saha_2000}. It was first discussed in \citet{1988ApJ...327..693G} from the perspective of lens equation, and \citet{Saha_2000} provided a comprehensive review of the degeneracy from the point of view of time delay surfaces. Here, we revisit this degeneracy in the context of GW lensing and formulate it using the dimensionless lens system. Because of the similarity transformation degeneracy, all (dark siren) strongly lensed GW observations will be unable to disentangle the size of the lens from the distances in the source-lens-observer system and cosmology. Complementary information (either from EM observations or otherwise) breaks the degeneracy.

We recall that the time delay of an image (relative to unlensed ray) is given by $$t_d(\vec{x}) = \frac{1+z_l}{c}\frac{D_l D_s}{D_{ls}} \theta_E^2\left[\frac{|\vec{x}-\vec{y}|^2}{2}-\Psi(\vec{x})\right],$$ which we emphasize again, $\vec{x}$ and $\vec{y}$ are dimensionless image and source position, and $\Psi(\vec{x})$ is the dimensionless deflection potential. Also, recall that the solutions of the dimensionless image position can be solved entirely from the dimensionless lens equation (Eq. (\ref{dimensionless lens eq})). Hence, one can write down the time delay in the following form:
\begin{equation}
    t_d(\vec{x}) = T_* \tau(\vec{x}),
    \label{t_d in T_* tau form}
\end{equation}
where we define $\tau(\vec{x})$ as the dimensionless (scale free) time delay:
\begin{equation}
    \tau(\vec{x})\equiv\left[\frac{|\vec{x}-\vec{y}|^2}{2}-\Psi(\vec{x})\right],
    \label{tau def}
\end{equation}
and $T_*$ as the time delay scaling:
\begin{equation}
    T_*\equiv\frac{1+z_l}{c}\frac{D_l D_s}{D_{ls}} \theta_E^2.
    \label{T_* def}
\end{equation}
Note that the time delay scaling $T_*$ is closely related to the time delay distance $D_{\Delta t}$ commonly used by other works by
\begin{equation}
    T_*=\frac{D_{\Delta t}}{c}\theta_E^2,
\end{equation}
as $D_{\Delta t}$ is given by
\begin{equation}
    D_{\Delta t}\equiv(1+z_l)\frac{D_l D_s}{D_{ls}}. 
\end{equation} 
Furthermore, assuming a flat universe (\textit{i.e.} $\Omega_k=0$) with fixed cosmology but a variable Hubble constant, we can see $\frac{D_l D_s}{D_{ls}}$ would be a function of source and lens redshift and the Hubble constant, as the angular diameter distance $D$ (at redshift $z$ away) is given by
\begin{equation}
    D(z,H_0) = \frac{c}{(1+z)H_0} \int_0^{z}\frac{dz'}{\sqrt{\Omega_r(1+z')^4+\Omega_m(1+z')^3+\Omega_\Lambda}},
\end{equation}
where $\Omega_r$, $\Omega_m$ and $\Omega_\Lambda$ are the (normalized) present-day radiation energy density, matter density, and dark energy density, respectively. Hence, the time delay scaling is a function of the lens and source redshift, the (angular) Einstein radius, as well as the Hubble constant: $T_*=T_*(z_l, z_s, \theta_E, H_0)$. All the information containing the scale of the system (\textit{i.e.} the redshifts, the angular diameter distances and the Einstein radius of the lens), as well as cosmological information (such as $H_0$) are coupled together in the time delay scaling $T_*$. Therefore, in GW lensing, when we measure the relative time delay between two images $i$ and $j$, $\Delta t_{d,ij}=T_*\tau_{ij}$, with $\tau_{ij}=(\tau_i-\tau_j)$ and $\tau_i=\tau(\vec{x}_i)$, we will have a common factor $T_*$ for all the image pairs. Since the factor $T_*$ enters as an overall multiplication in the time delay, one cannot disentangle the angular diameter distances, the Einstein radius $\theta_E$ and the Hubble constant $H_0$ from each other, as they are coupled together inside $T_*$. In other words, different sets of $\{z_l, z_s, \theta_E, H_0\}$ can give the same value of $T_*$, and we would not be able to separate the two by just examining $\Delta t_{d,ij}$.

Meanwhile, in GW lensing, we also measure the effective luminosity distance of each image $d_{L,i}^\text{eff}={d_L}/{\sqrt{|\mu_i|}}$.\footnote{We stress again that $\mu_i$ is scale-free.} We then also note that the original luminosity distance would be a function of source redshift and Hubble constant: $d_L=d_L(z_s, H_0)$, as 
\begin{equation}
    d_L=(1+z_s)\frac{c}{H_0} \int_0^{z_s}\frac{dz'}{\sqrt{\Omega_r(1+z')^4+\Omega_m(1+z')^3+\Omega_\Lambda}}.
    \label{d_L eq}
\end{equation} 
We note that in GW observation of compact binary coalescence, we cannot measure $z_s$ directly due to the source mass-redshift degeneracy.\footnote{It may be possible for binary neutron star merger if the equation of state is known and tidal effects are analyzed, but we neglect these indirect methods in the following analysis \citep{Messenger_2014, Pang_2020}.} The role of $d_L$ is very similar to $T_*$: for all the images and their observed effective luminosity distances, $d_L$ would appear as a common factor, and both the redshift $z_s$ and the Hubble constant $H_0$ are coupled inside $d_L$. Therefore, the redshifts and the cosmology are typically inferred using the directly measurable GW parameters.

\begin{table*}
\centering
\begin{tabular}{|c | c | c | c|} 
 \hline
 Type of parameters & $\vec{\Phi}_l$ (SIE lens) & $\vec{\Phi}_s$ (GW source) & $\vec{\Phi}_{c}$ (cosmology) \\ [0.5ex] 
 \hline
 Setup 1 & $z_l=0.4, v=6 \times 10^{-4} c, f = 0.7$ & $z_s=1.8, \vec{\beta}\approx(4.45\times 10^{-3}, 6.35\times 10^{-3}) \text{ (in arcseconds)}$ & $H_0=67.8 \text{ km} \text{ s}^{-1} \text{ Mpc}^{-1}$ \\
  \hline
Setup 2 & $z_l=0.8, v=5.83 \times 10^{-4} c, f = 0.7$ & $z_s=1.8, \vec{\beta}\approx(2.63\times 10^{-3}, 3.76\times 10^{-3}) \text{ (in arcseconds)}$ & $H_0=67.8 \text{ km} \text{ s}^{-1} \text{ Mpc}^{-1}$ \\
 \hline
 Setup 3 & $z_l=1.1, v=6.18 \times 10^{-4} c, f = 0.7$ & $z_s=1.7, \vec{\beta}\approx(1.72\times 10^{-3}, 2.45 \times 10^{-3}) \text{ (in arcseconds)}$ & $H_0=63.2 \text{ km} \text{ s}^{-1} \text{ Mpc}^{-1}$ \\
 \hline
\end{tabular}
\begin{tabular}{|c | c | c|} 
 \hline
 Lensing observables & $d_{L,1}^\text{eff}, d_{L,2}^\text{eff}, d_{L,3}^\text{eff}, d_{L,4}^\text{eff}$ (in Mpc) & $\Delta t_{d,12}, \Delta t_{d,23}, \Delta t_{d,34}$ (in seconds)\\ [0.5ex]
 \hline
  Common result for setup 1, 2, 3   & $6383, 6216, 7260, 7624$ & $-33629, -228960, -42685$\\
 \hline
\end{tabular}
\caption{Three different lens system setups leading to the same lensing observables under similarity transformation degeneracy: for the SIE lens parameters, $z_l, v, f$ are the redshift, velocity dispersion and SIE axis ratio respectively and $c$ is the speed of light. For the GW source, $z_s, \vec{\beta}$ are the redshift and the (angular) source position in arcseconds respectively. Lastly, $H_0$ is the Hubble constant. The first and second setup share the same source redshift and Hubble constant, but the second setup has a larger lens redshift and smaller velocity dispersion compared to the first. Meanwhile, the first setup has relatively low lens redshift and also a smaller lens galaxy velocity dispersion $v$ compared to the third setup. For the angular Einstein radius, the first setup has a larger angular Einstein radius of $0.635$ arcseconds versus the second setup's $0.376$ arcseconds and the third setup's $0.245$ arcseconds. Since the three systems' dimensionless system are the same (\textit{i.e.} they share the same dimensionless source position $\vec{y}$ and SIE axis ratio $f$), and they share the same value of time delay scaling $T_*$ and luminosity distance $d_L$, so they share the same observed effective luminosity distances $d_{L,i}^\text{eff}$ and Morse index for each of the four images (with the first two images having $n=0$ and the last two images having $n=1$), as well as the relative time delay between the image pairs $\Delta t_{d,ij}$ as shown above. The physical picture of the degeneracy represented by the first two setups is illustrated in Fig. \ref{two lens system under similarity transformation}.}
\label{table: STD 2 degenerate setup}
\end{table*}

\begin{figure}
\centering
\includegraphics[scale=0.38]{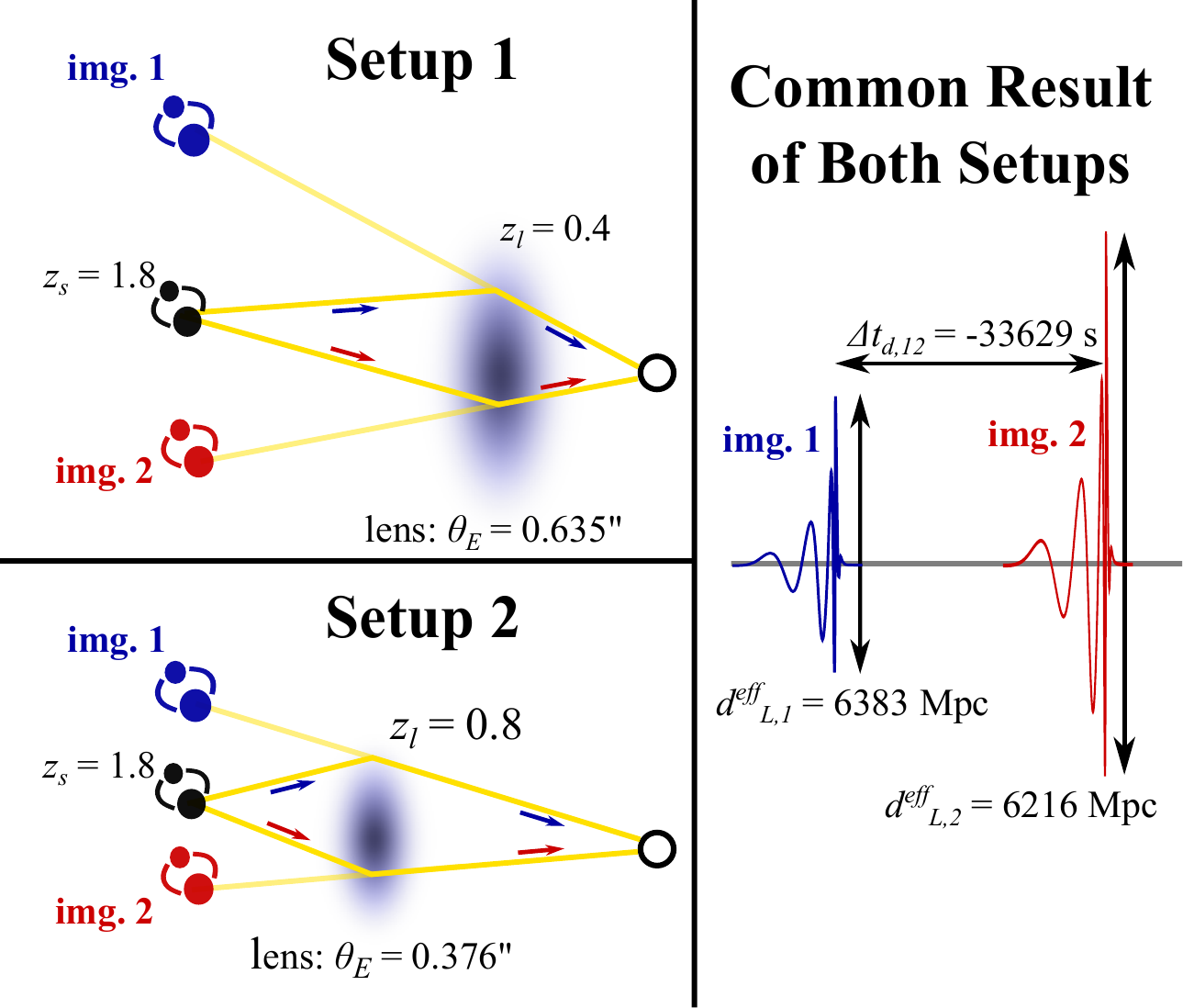}
\caption{Illustrative example of similarity transformation degeneracy in GW lensing: the upper and lower lens system on the left panel corresponds to Setup 1 and 2 described in Table \ref{table: STD 2 degenerate setup} respectively. They share the same dimensionless system, source redshift and Hubble constant, but they have different lens redshift $z_l$ ($0.4$ versus $0.8$), velocity dispersion $v$ ($6 \times 10^{-4} c$ versus $5.83 \times 10^{-4} c$) and hence Einstein radius $\theta_E$ ($0.635$ arcseconds versus $0.376$ arcseconds). However, as shown on the right panel and in Table \ref{table: STD 2 degenerate setup}, they lead to the same measurement of relative time delays and effective luminosity distances of the images as their dimensionless system are the same and the values of $T_*$ and $d_L$ also match.}
\label{two lens system under similarity transformation}
\centering
\end{figure}

\subsection{Degenerate transformations} \label{S4.1}

\subsubsection{$z_l$-$\theta_E$ degeneracy}
Let us illustrate the degeneracy between the distance and the size of the lens. Suppose, that we introduce a transformation such that the distance to lens changes in such a way that the overall time-delay distance changes by factor $a$. Suppose, furthermore, that the Einstein radius of the lens changes by factor $1/a^{1/2}$ with all other quantities kept constant.\footnote{The dimensionless source/image positions are kept fixed. Because the lens equation only depends on these dimensionless quantities, they are not affected by the transformation. } That is, we introduce the following transformation: 
\begin{equation*}
    \begin{split}
        &z_l\rightarrow z_l^\prime=x_0 z_l\,,\\
        &\theta_E\rightarrow \theta_E^\prime=\theta_E/a^{1/2}\,,\\
        &D_{\Delta t}\rightarrow D_{\Delta t}^\prime = D_{\Delta t} a\,.
    \end{split}
\end{equation*}
(Here $x_0$ is chosen such that the Einstein radius gains a factor $a^{-1/2}$.) 
In this case, the overall time delay does not change:
\begin{equation*}
    T_*\rightarrow T_*^\prime=\frac{D_{\Delta t}^\prime}{c}(\theta_E^\prime)^2=\frac{D_{\Delta t}}{c}\theta_E^2=T_*
\end{equation*}
Since the magnification depends only on the dimensionless quantities, they do not change. For this reason, the effective luminosity distances also do not change. Thus, under the transformation, all GW lensing observables are unchanged:
\begin{equation}
\begin{split}
    &d_{L,i}^\text{eff} \rightarrow d_{L,i}^{\text{eff},\prime} = d_{L,i}^{\text{eff}} \\
    &\Delta t_{d,ij} \rightarrow \Delta t_{d,ij}^\prime = \Delta t_{d,ij}\,, \\
    &n_j\rightarrow n_j^\prime=n_j\,.
\end{split}
\end{equation}
That is, the lens system before/after produces identical GW observables. 

\subsubsection{$H_0$-$\theta_E$ degeneracy}
Alternatively, one can introduce a transformation in the Hubble constant and the angular size of the lens (all other quantities kept constant):
\begin{equation*}
    \begin{split}
        &H_0\rightarrow H_0^\prime=H_0/a \,,\\
        &\theta_E\rightarrow \theta_E^\prime=\theta_E/a^{1/2}\,,\\
        &D_{\Delta t}\rightarrow D_{\Delta t}^\prime = D_{\Delta t} a\,,
    \end{split}
\end{equation*}
which is likewise degenerate since 
\begin{equation}
\begin{split}
    &d_{L,i}^\text{eff} \rightarrow d_{L,i}^{\text{eff},\prime} = d_{L,i}^{\text{eff}} \\
    &\Delta t_{d,ij} \rightarrow \Delta t_{d,ij}^\prime = \Delta t_{d,ij}\,, \\
    &n_j\rightarrow n_j^\prime=n_j\,.
\end{split}
\end{equation}

\subsubsection{Partial $z_s$-$z_l$-$\theta_E$-$H_0$ degeneracy}
Finally, there exists a non-trivial transformation involving the source redshift transformations, the Hubble constant, the scale of the lens, when keeping the luminosity distance fixed: 
\begin{equation*}
    \begin{split}
        &z_s\rightarrow z_s^\prime = x_1 z_s\,,\\
        &H_0\rightarrow H_0^\prime=H_0/a \,,\\
        &d_L\rightarrow d_L^\prime=d_L(z_s^\prime,H_0^\prime)=d_L \\
        &\theta_E\rightarrow \theta_E^\prime=\theta_E/b^{1/2}\,,\\
        &D_{\Delta t}\rightarrow D_{\Delta t}^\prime = D_{\Delta t} b\,,
    \end{split}
\end{equation*}
where $x_1$ is chosen such that $d_L^\prime=d_L$. 
The above transformation likewise leads to the same GW observables: 
\begin{equation}
\begin{split}
    &d_{L,i}^\text{eff} \rightarrow d_{L,i}^{\text{eff},\prime} = d_{L,i}^{\text{eff}} \\
    &\Delta t_{d,ij} \rightarrow \Delta t_{d,ij}^\prime = \Delta t_{d,ij}\,, \\
    &n_j\rightarrow n_j^\prime=n_j\,.
\end{split}
\end{equation}
However, it is worth noting that this final degeneracy is a \emph{partial} degeneracy, because $d_L$ must be kept fixed in the transformation. 
For this reason, the degeneracy can be broken by measuring $z_l$ and $\theta_E$ only, even without $z_s$ measurement. \footnote{More technically, if there are $4$ degrees of freedom in allowing for $z_s$, $z_l$, $\theta_E$, and $H_0$ to vary, then the condition $d_L^\prime=d_L$ produces one "constraint", such that there are really only $4-1=3$ degrees of freedom, so a measurement of two out of the four degenerate quantities is sufficient to break the degeneracy.} Section \ref{S4.4} will provide a more detailed formalism for this partial degeneracy.

\subsubsection{What quantities are degenerate?}
From the above considerations, one should conclude that it is possible that for two systems with very different absolute scale and distances as well as different value of the Hubble constant (\textit{i.e.} different set of $\{z_l, z_s, \theta_E, H_0\}$).
That is, the scale of the lens $\theta_E$, the redshift to the lens and source $z_l, z_s$, and the Hubble constant $H_0$ are entirely degenerate with one another. To further illustrate the effect, Table \ref{table: STD 2 degenerate setup} shows three very different SIE lens systems that would lead to the same lensing observables. The first two systems considered are illustrated in Fig. \ref{two lens system under similarity transformation}.
\citet{Saha_2000} referred to the transformation that induced the degeneracy presented above as the similarity transformation; therefore, we call this degeneracy the similarity transformation degeneracy.
To back up these arguments, we will give further evidence of these degeneracies in our lens reconstruction results, where the degeneracies between $\{z_l, z_s, \theta_E, H_0\}$ are obvious.

\subsection{Lens Reconstruction with Similarity Transformation Degeneracy}\label{S4.2}
What can GWs tell us about the galaxies that lens them? For dark siren GW lensing (\textit{i.e.} observation without EM counterpart or any other supplementary information), we can only measure the relative time delays $\Delta t_{d,ij}$, effective luminosity distances $d_{L,i}^\text{eff}$ and relative Morse indices $\Delta n_{ij}$ of the images \footnote{The absolute Morse phase of type-II images may also be measurable when the GW experiences higher-order-modes \citep{dai2017waveforms, Ezquiaga_2021, Janquart_2021b, Wang_2021a}.}. In this case, only a limited amount of information can be reconstructed. We focus on galaxy lenses here: as the image with Morse index $n_j = 2$ (\textit{i.e.} the type III image) is a highly demagnified (central) image that can rarely be observed in galaxy lensing \citep{2013ApJ...773..146D, 2016MNRAS.456.2210C, 2017ApJ...843..148C}, we neglect it. Meanwhile, the images with Morse index $n_j = \{0, 1\}$ correspond to positive and negative value of magnification $\mu$ respectively \citep{vijaykumar2023detection}. In addition, for a typical galaxy lens system with four images, the arrival order of images is type I-I-II-II \citep{1992grle.book.....S, Wierda_2021}. Hence, assuming all four images are observed, we can sort out the image types by arrival order, which would be included in the lens reconstruction process. GWs can thus yield the effective luminosity distances $d_{L,i}^\text{eff}$, relative time delays $\Delta t_{d,ij}$, and absolute Morse phases $n_i$ as observables. However, it is important to understand what lensing quantities these observables can map to; it is clear that due to the similarity transformation degeneracy, not all lensing quantities can be measured.\footnote{For example, can we retrieve the Einstein radius? How about the axis ratio? What about the redshifts?}

From a reverse engineering point of view, after we obtain the GW observables ($d_{L,i}^\text{eff}, \Delta t_{d,ij}, n_i$), we can obtain the absolute value of the magnification ratio of images by
\begin{equation}
    \left|\frac{\mu_i}{\mu_j}\right| = \left(\frac{d_{L,j}^\text{eff}}{d_{L,i}^\text{eff}}\right)^2,
\end{equation}
which is completely determined by the dimensionless lens system\footnote{That is, it only depends on the scaled image/source positions and the scaled lens potential, but not on the angular diameter distances or the Einstein radius (scale of the lens).} after $d_L$ is cancelled out. We note that, for our galaxy lensing case where we ignore type III images, one can also obtain the magnification ratio (signs included) by taking the Morse index into account:
\begin{equation}
    \frac{\mu_i}{\mu_j} = \left(\frac{d_{L,j}^\text{eff}}{d_{L,i}^\text{eff}}\right)^2 (-1)^{\Delta n_{ij}}.
\end{equation}

Meanwhile, we can also make use of the ratio of relative time delay of different image pairs and obtain the ratio of dimensionless time delay:
\begin{equation}
     \frac{\tau_{ij}}{\tau_{mn}} = \frac{\Delta t_{d,ij}}{\Delta t_{d,mn}},
\end{equation}
which is completely fixed by the dimensionless lens system, meaning it does not depend on the time-delay scaling $T_*$. Suppose we have sufficient number of observed images and a parametric lens model (for instance, four images and an SIE lens mass profile). In that case, we can reconstruct the dimensionless lens system from the effective luminosity distance ratios and relative time delay ratios. For instance, if we choose SIE to be our lens model, using the magnification ratios ${\mu_i}/{\mu_j}$ and dimensionless time delay ratios ${\tau_{ij}}/{\tau_{mn}}$, one would be able to reconstruct the axis ratio $f$ (which is part of the dimensionless system), as well as the dimensionless source position $\vec{y}$ (hence also the dimensionless image positions $\vec{x}$ by Eq. (\ref{dimensionless lens eq})). In Section \ref{S4.5}, we use a SIS lens to demonstrate this idea analytically.

After reconstructing the dimensionless lens system, using Eq. (\ref{tau def}), we can calculate the dimensionless time delay $\tau_i$ of the i-th image. Then we can obtain the value of $T_*$ from the observed $\Delta t_{d,ij}$ using Eq. (\ref{t_d in T_* tau form}). Similarly, we can calculate the i-th image's magnification $\mu_i$ from Eq. (\ref{magnification eq}) and use the relation $d_L=d_{L,i}^\text{eff}\sqrt{|\mu_i|}$ to calculate $d_L$.

What, therefore, can GWs alone infer? GWs can be used to recover the dimensionless system: dimensionless source position $\vec{y}$ and dimensionless deflection potential parameters $\vec{\Phi}_\Psi$ from the ratios ${\mu_i}/{\mu_j}$ and ${\tau_{ij}}/{\tau_{mn}}$. However, in terms of the scale of the system, we can only recover the time-delay factor/scaling $T_*$ and the absolute luminosity distance $d_L$. Therefore, we will not be able to infer the actual physical scales and sizes (such as $D_s$, $D_l$ and $\theta_E$) and the actual mass distribution $\Sigma$. In other words, we can only reconstruct the posterior $P(\vec{y}, \vec{\Phi}_\Psi, T_*, d_L|\vec{d}_{GW})$ (under a selected lens model and form of $\Psi(\vec{x})$), but not the posterior of all free parameters $P(\vec{\Phi}_{all}|\vec{d}_{GW})$ that would allow inferring the cosmological parameters from the lens system. 
Therefore, in the process of lens reconstruction from dark siren lensed GW in the Bayesian framework, we should only set $\{\vec{y}, \vec{\Phi}_\Psi, T_*, d_L\}$ as the (irreducible) free parameters in our lens model, instead of $\vec{\Phi}_{all}=(\vec{\beta}, z_s, \theta_E,\vec{\Phi}_\Psi, z_l, H_0)$. That is, the inference of $\{\vec{y}, \vec{\Phi}_\Psi, T_*, d_L\}$ is a data-driven, while inference of $\vec{\Phi}_{all}$ is a prior-driven.\footnote{In particular, prior information about $z_l$ and $\theta_E$ are required to perform meaningful reconstructions with $\vec{\Phi}_{all}$ (such a prior-information-driven approach was demonstrated in \citet{Jana_2023}, which supports the assertion).}

\begin{table*}
\centering
\begin{tabular}{|c | c | c | c|} 
 \hline
 Type of parameters & $\vec{\Phi}_l$ (SIE lens) & $\vec{\Phi}_s$ (GW source) & $\vec{\Phi}_{c}$ (cosmology) \\ [0.5ex] 
 \hline
  Injection values & $z_l=0.5, v=0.0006c, f = 0.5$ & $z_s=2, \vec{\beta}\approx(0.00592,0.00887) \text{ (in arcseconds)}$ & $H_0=67.8 \text{ km} \text{ s}^{-1} \text{ Mpc}^{-1}$ \\
 \hline
\end{tabular}
\caption{Injection values of the system parameters for our lens reconstruction example: 1. for the lens parameters, $z_l, v, f$ are the redshift, velocity dispersion and SIE axis ratio, respectively. Note that $c$ is the speed of light and we align the long axis of the SIE with the vertical axis (\textit{i.e.} the $y_2$ or $x_2$ axis). 2. For the source, $z_s, \vec{\beta}$ are the redshift and the (angular) source position in arcseconds, respectively. 3. For cosmology, $H_0$ is the Hubble constant.}
\label{table:dark_siren_lens_reconstruction_injection_setup}
\end{table*}

\begin{table*}
\centering
\begin{tabular}{|c | c | c|} 
 \hline
 Observables & Effective luminosity distance: $d_{L,1}^\text{eff}, d_{L,2}^\text{eff}, d_{L,3}^\text{eff}, d_{L,4}^\text{eff}$ (in Mpc) & Relative time delay between images: $\Delta t_{d,12}, \Delta t_{d,23}, \Delta t_{d,34}$ (in s) \\ [0.5ex] 
 \hline
 Injected values & $9404,9237,12596,13169$ & $-55817,-491735,-66594$\\
 \hline
\end{tabular}
\caption{GW lensing observables of injected system setup shown in Table \ref{table:dark_siren_lens_reconstruction_injection_setup}. Note that we have four images, and the image types are I-I-II-II.}
\label{table:dark_siren_lens_reconstruction_injected_observables}
\end{table*}

\begin{figure}
\centering
\includegraphics[scale=0.45]{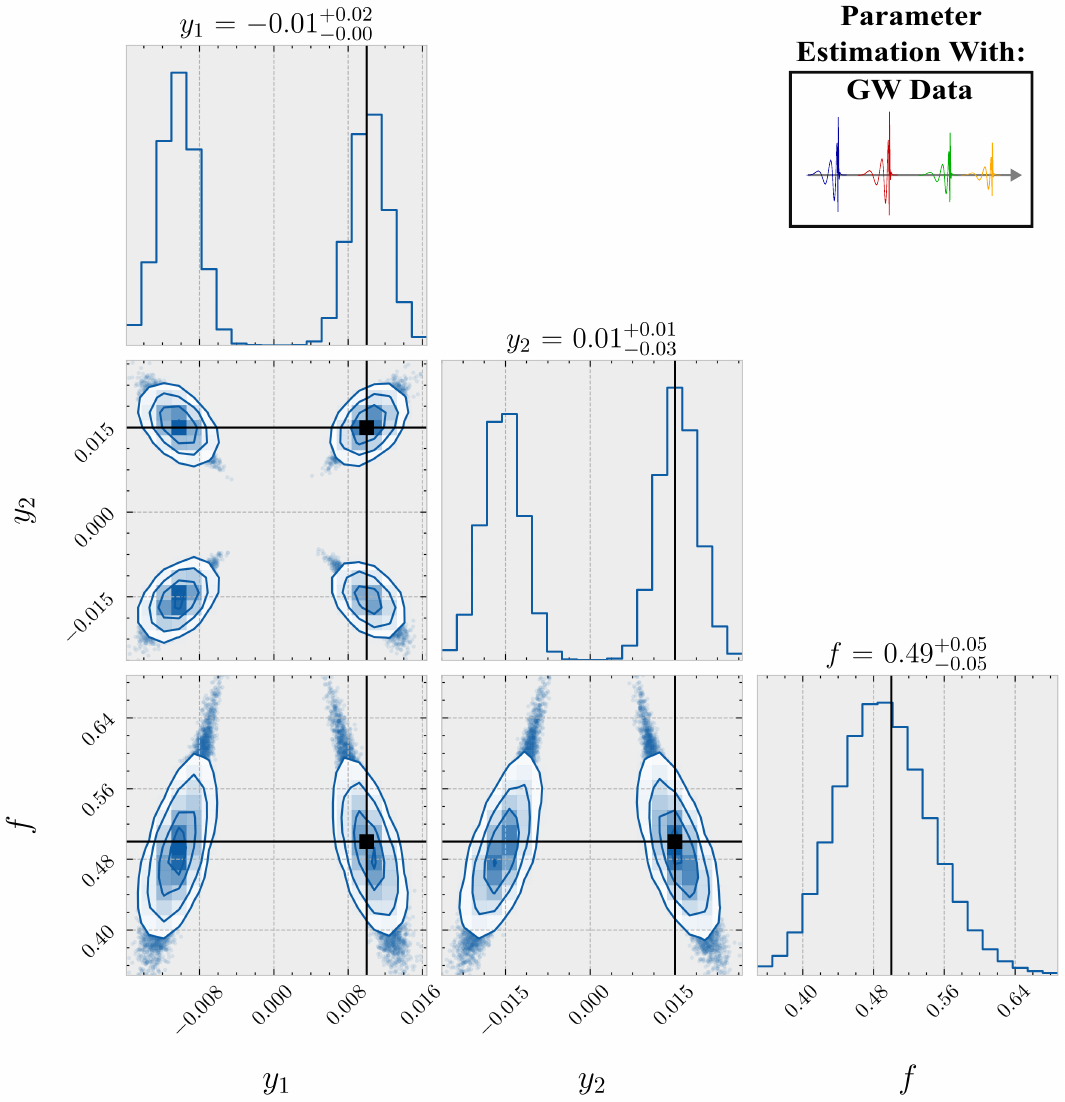}
\caption{Result of SIE lens reconstruction of dark siren lensed GW using injected parameters shown in Table \ref{table:dark_siren_lens_reconstruction_injection_setup} and Table \ref{table:dark_siren_lens_reconstruction_injected_observables}, which an uncorrelated Gaussian likelihood with $5\%$ error on observed relative time delays and effective luminosity distances is assumed. After taking similarity transformation degeneracy into account, the five free parameters in the inference are $\{\vec{y}=(y_1, y_2), f, T_*, d_L\}$, which are the dimensionless source position, the SIE axis ratio, the time delay scaling $T_*\equiv\frac{1+z_l}{c}\frac{D_l D_s}{D_{ls}} \theta_E^2$ and the (unlensed) luminosity distance respectively. In this figure, we show the posterior distribution of the parameters that represent the dimensionless lens system: $\vec{y}=(y_1, y_2)$ and $f$. We can see, the parameters are recovered fairly well. We note that the dimensionless source position suffers from multiple solutions only because SIE lens has two degrees of reflectional symmetry. Other than that, there is no degeneracy among the parameters.
}
\label{fig:STD_source_vs_f}
\centering
\end{figure}

\begin{figure}
\centering
\includegraphics[scale=0.62]{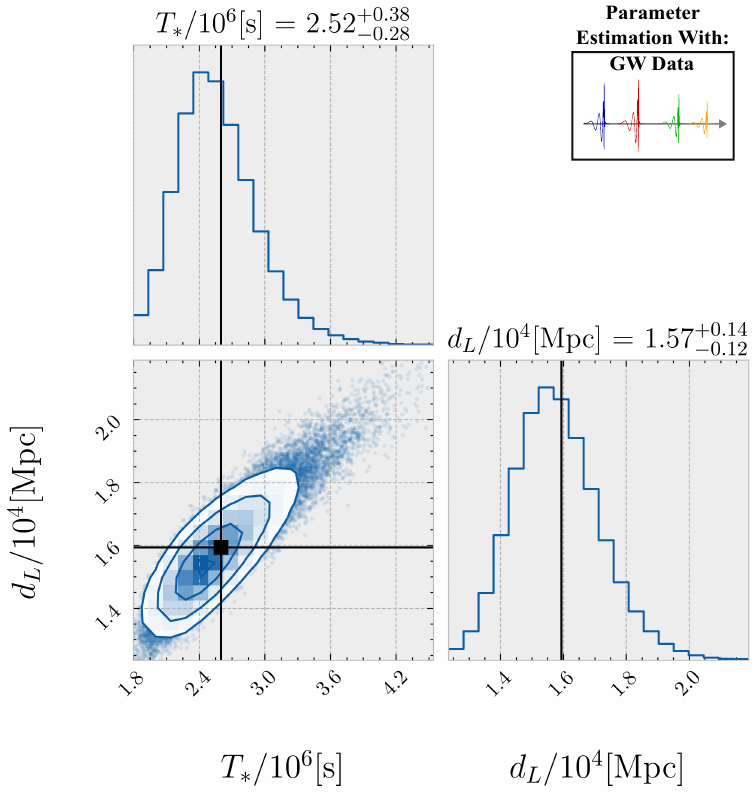}
\caption{Result of SIE lens reconstruction of dark siren lensed GW using injected parameters shown in Table \ref{table:dark_siren_lens_reconstruction_injection_setup} and Table \ref{table:dark_siren_lens_reconstruction_injected_observables}, which an uncorrelated Gaussian likelihood with $5\%$ error on observed relative time delays and effective luminosity distances is assumed. After taking similarity transformation degeneracy into account, the five free parameters are $\{\vec{y}=(y_1, y_2), f, T_*, d_L\}$, which are the dimensionless source position, the SIE axis ratio, the time delay scaling $T_*\equiv\frac{1+z_l}{c}\frac{D_l D_s}{D_{ls}} \theta_E^2$ and the (unlensed) luminosity distance respectively. In this figure, we show the posterior distribution of the parameters that represent the scale of the lens system: $T_* \text{ (in seconds)}$ and $d_L \text{ (in Mpc)}$. We can see, the two parameters are recovered fairly well.
}
\label{fig:STD_T_vs_dL}
\centering
\end{figure}

Below, we demonstrate parameter estimation results using our "irreducible" parameters, which accounts for the similarity transformation degeneracy. We inject an SIE lens with the parameters described in Table \ref{table:dark_siren_lens_reconstruction_injection_setup}. After obtaining the injected observables for the four images as described in Table \ref{table:dark_siren_lens_reconstruction_injected_observables}, we perform Bayesian parameter estimation of the lens system. For the likelihood $P(\vec{d}_{GW}|\vec{\nu}_{im})$, we assume a simple and uncorrelated 5\% Gaussian error on the measured $d_{L,j}^\text{eff}$ and $\Delta t_{d,ij}$\footnote{This is to roughly account for both measurement and lens model uncertainties. Note that a correlated probability distribution of the measured $d_{L,j}^\text{eff}$ and $\Delta t_{d,ij}$ that emerge from direct GW data analysis (such as \citet{haris2018identifying, Hannuksela_2019, Goyal_2021a, Janquart_2021, janquart2022golum, Janquart_2023a, Janquart_2023, Lo_2023, 2023arXiv230408393T}) should be used in a quantitative analysis; we are primarily interested in understanding the various lensing degeneracies qualitatively, where a rough representative error estimate should be sufficient.} Under the similarity transformation degeneracy for dark siren GW lensing, we would only sample five free parameters in this case: the two-dimensional dimensionless source position $\vec{y}=(y_1, y_2)$, the SIE axis ratio $f$, the luminosity distance $d_L$ and the time delay scaling $T_*$. We also assume a uniform prior for all parameters for simplicity. Fig.~\ref{fig:STD_source_vs_f} and Fig.~\ref{fig:STD_T_vs_dL} show the lens reconstruction result of the system's dimensionless parameters ($\vec{y}, f$) and parameters that represent the scale of the system ($T_*, d_L$) respectively. Firstly, there are four peaks in the posterior for the dimensionless source position $\vec{y}$. This is to be expected due to the reflectional symmetry of the SIE lens. Then, we can see for all five parameters, the injected values are included in the recovered posteriors. The full parameter estimation result of all five parameters is shown in Fig.~\ref{fig:STD_standard_PE} in the Appendix.

\begin{figure*}
\centering
\includegraphics[scale=0.5]{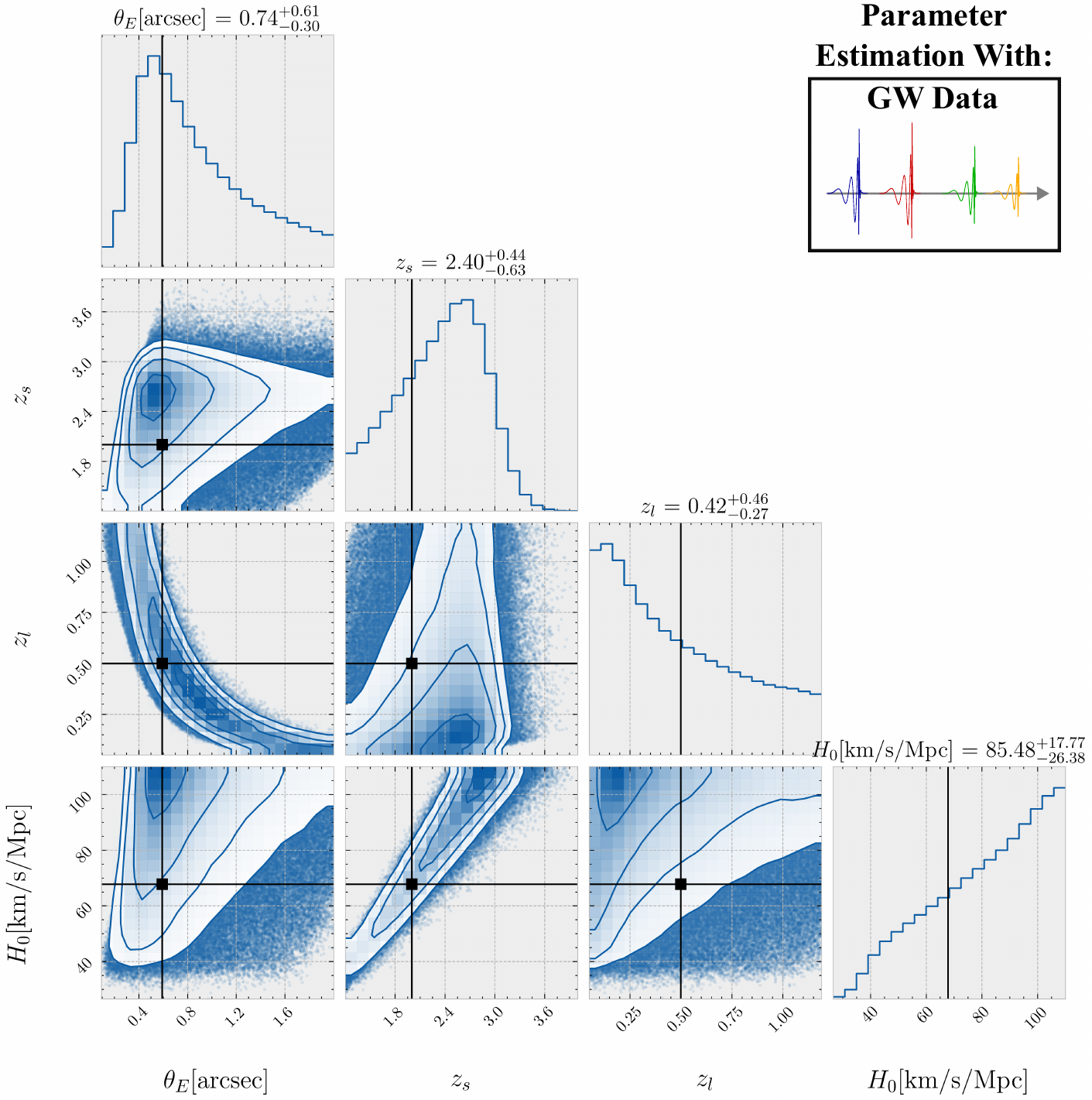}
\caption{Result of SIE lens reconstruction of dark siren lensed GW using injected parameters shown in Table \ref{table:dark_siren_lens_reconstruction_injection_setup} and Table \ref{table:dark_siren_lens_reconstruction_injected_observables}, without accounting for similarity transformation degeneracy. An uncorrelated Gaussian likelihood with $5\%$ error on observed relative time delays and effective luminosity distances is also assumed. The seven free parameters are $\{\vec{y}=(y_1, y_2), f, \theta_E \text{ (in arcseconds)}, z_s, z_l, H_0 \text{ (in km} \text{ s}^{-1} \text{ Mpc}^{-1}\text{)}\}$, which are the dimensionless source position, the SIE axis ratio, the Einstein radius, the source and lens redshift and the Hubble constant respectively. Here we show the posterior distribution of the four parameters that represent the scale of the system: $\theta_E \text{ (in arcseconds)}, z_s, z_l, H_0 \text{ (in km} \text{ s}^{-1} \text{ Mpc}^{-1}\text{)}$. They are poorly recovered (which can be sensitive to prior choice) and showed high degree of degeneracies, which is expected from similarity transformation degeneracy.}
\label{fig:STD_degenerate_case_einR_vs_zs_vs_zl_vs_H0}
\centering
\end{figure*}

We then repeat the exercise with 7 free parameters: $\{\vec{y}=(y_1, y_2), f, \theta_E, z_s, z_l, H_0\}$, which does not account for similarity transformation degeneracy. The setup is identical to the run presented in Fig.~\ref{fig:STD_source_vs_f} and Fig.~\ref{fig:STD_T_vs_dL}, with a uniform prior on all sampling parameters. Fig.~\ref{fig:STD_degenerate_case_einR_vs_zs_vs_zl_vs_H0} shows the posterior distribution of the parameters that describe the scale of the system: namely the Einstein radius $\theta_E$, the source and lens redshifts $z_s, z_l$ and the Hubble constant $H_0$. The full result of the parameter estimation is shown in Fig.~\ref{fig:STD_degenerate_case_full_PE} in Appendix. As expected, the four quantities that describe the actual scale of the system, $\{\theta_E, z_s, z_l, H_0\}$, are poorly recovered and constrained due to the degeneracies that are present among these parameters: for instance, the $z_s\text{-}H_0$ relation is spanned across the entire parameter space diagonally, implying that the two are fully degenerate (due to their combination within $d_L$). Similarly, the $z_l\text{-}\theta_E$ pair is also highly degenerate. Meanwhile, we note that the dimensionless system parameters are still recovered as shown in Fig. \ref{fig:STD_degenerate_case_full_PE}. Therefore, despite having 7 constraints from the dark siren GW observations (4 $d_{L,j}^\text{eff}$ and 3 $\Delta t_{d,ij}$), that, naively, could be matched to the 7 free lensing parameters ($f$, $v$, $\vec{y}$, $z_s$, $z_s$, $H_0$), one cannot constrain the full system due to the similarity transformation degeneracy. Indeed, these results support our earlier discussion.

A detected, strongly lensed GWs would give us a measurement of the relative time delays $\Delta t_{d,ij}$ and effective luminosity distances $d_{L,i}^\text{eff}$. With just $P(\Delta t_{d,ij})$ and $P(d_{L,i}^\text{eff})$, as demonstrated above, we would suffer from similarity transformation degeneracies and can only infer a reduced number of parameters of the system. This may not be very useful for science application that rely on lens reconstruction. However, if the lens system is localised through an archival search or a rapid EM follow-up, complementary EM information of the corresponding lens system, especially the lens and source redshift and Einstein radius of the lens, can break the similarity transformation degeneracy. In Section \ref{S4.4}, we demonstrate that such identification and EM information is crucial to the Hubble constant measurement.

\subsection{Lens Reconstruction with Two or Three Images Only}\label{S4.3}
Here, we provide additional examples of dark siren GW SIE lens reconstruction attempt with only two or three of all four GW images are used in the process of parameter estimation. In practice, this may have to be done when the remaining lensed images have not arrived yet, or they are too dim to be detected. We also assume the order of the two or three images are identified or assumed correctly within all images. \footnote{To elaborate, we first obtain the injected observables from, for instance, the first three images (in practice, since the first two images are type I image and the third one is type II image, we would not be able to distinguish if we observed the first three images or the first two and fourth images from the lens system). Then we attempt lens reconstruction by (correctly) assuming that the three images we obtained are the first three of the four images.} The setup of the lens system is presented in Table \ref{table:dark_siren_lens_reconstruction_injection_setup} and Table \ref{table:dark_siren_lens_reconstruction_injected_observables}.

\begin{figure}
\centering
\includegraphics[scale=0.45]{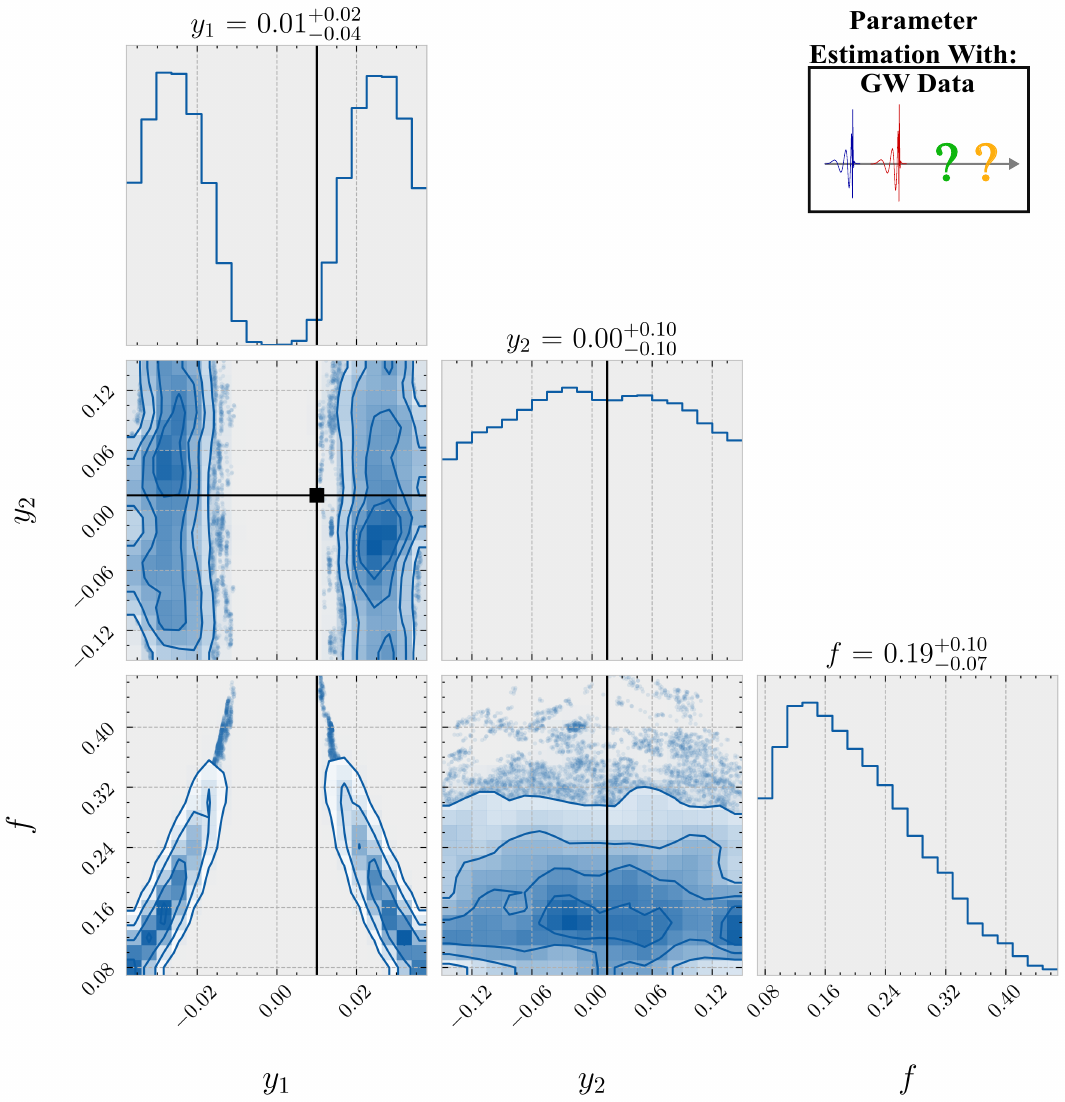}
\caption{Result of SIE lens reconstruction of dark siren lensed GW using injected parameters shown in Table \ref{table:dark_siren_lens_reconstruction_injection_setup} and the observables of the first two images from Table \ref{table:dark_siren_lens_reconstruction_injected_observables}, which an uncorrelated Gaussian likelihood with $0.001\%$ error on observed relative time delays and effective luminosity distances is assumed. After taking similarity transformation degeneracy into account, the five free parameters in the inference are $\{\vec{y}=(y_1, y_2), f, T_*, d_L\}$, which are the dimensionless source position, the SIE axis ratio, the time delay scaling $T_*\equiv\frac{1+z_l}{c}\frac{D_l D_s}{D_{ls}} \theta_E^2$ and the (unlensed) luminosity distance respectively. In this figure, we show the posterior distribution of the parameters that represent the dimensionless lens system: $\vec{y}=(y_1, y_2)$ and $f$. We can see, the parameters cannot be recovered and the posterior is uninformative.
}
\label{STD 2 img source vs f}
\centering
\end{figure}

Fig. \ref{STD 2 img source vs f} presented the result of parameter estimation of the dimensionless system parameters $\vec{y}=(y_1, y_2)$ and $f$ in the case of only using the first two images. The full result is shown in Fig. \ref{STD 2 img standard PE} in the Appendix. We note that an uncorrelated Gaussian likelihood with only $0.001\%$ error on the observables is assumed in order to minimize uncertainty. As shown, the injected parameters are not recovered and the posteriors appears uninformative. This is expected, as there is a mismatch in degree of freedom: we only have three observables ($d_{L,1}^\text{eff}, d_{L,2}^\text{eff}$ and $\Delta t_{d,12}$), but the entire lens system has five irreducible free parameters ($\{\vec{y}=(y_1, y_2), f, T_*, d_L\}$).

\begin{figure}
\centering
\includegraphics[scale=0.45]{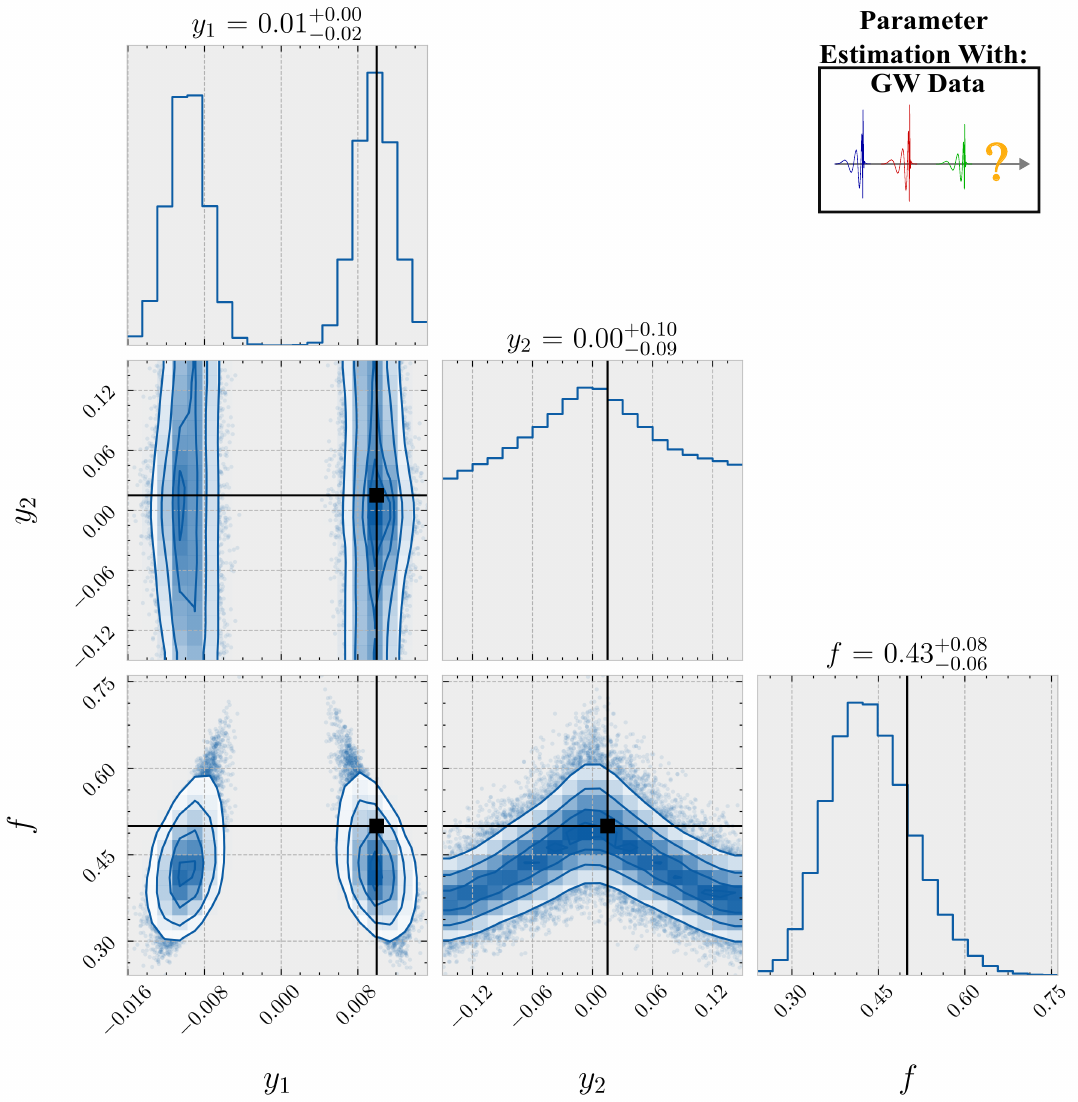}
\caption{Result of SIE lens reconstruction of dark siren lensed GW using injected parameters shown in Table \ref{table:dark_siren_lens_reconstruction_injection_setup} and the observables of the first three images from Table \ref{table:dark_siren_lens_reconstruction_injected_observables}, which an uncorrelated Gaussian likelihood with $5\%$ error on observed relative time delays and effective luminosity distances is assumed. After taking similarity transformation degeneracy into account, the five free parameters in the inference are $\{\vec{y}=(y_1, y_2), f, T_*, d_L\}$, which are the dimensionless source position, the SIE axis ratio, the time delay scaling $T_*\equiv\frac{1+z_l}{c}\frac{D_l D_s}{D_{ls}} \theta_E^2$ and the (unlensed) luminosity distance respectively. In this figure, we show the posterior distribution of the parameters that represent the dimensionless lens system: $\vec{y}=(y_1, y_2)$ and $f$. We can see, the parameters can be recovered, but with very large uncertainty. 
}
\label{STD 3 img source vs f (5percent error)}
\centering
\end{figure}

\begin{figure}
\centering
\includegraphics[scale=0.45]{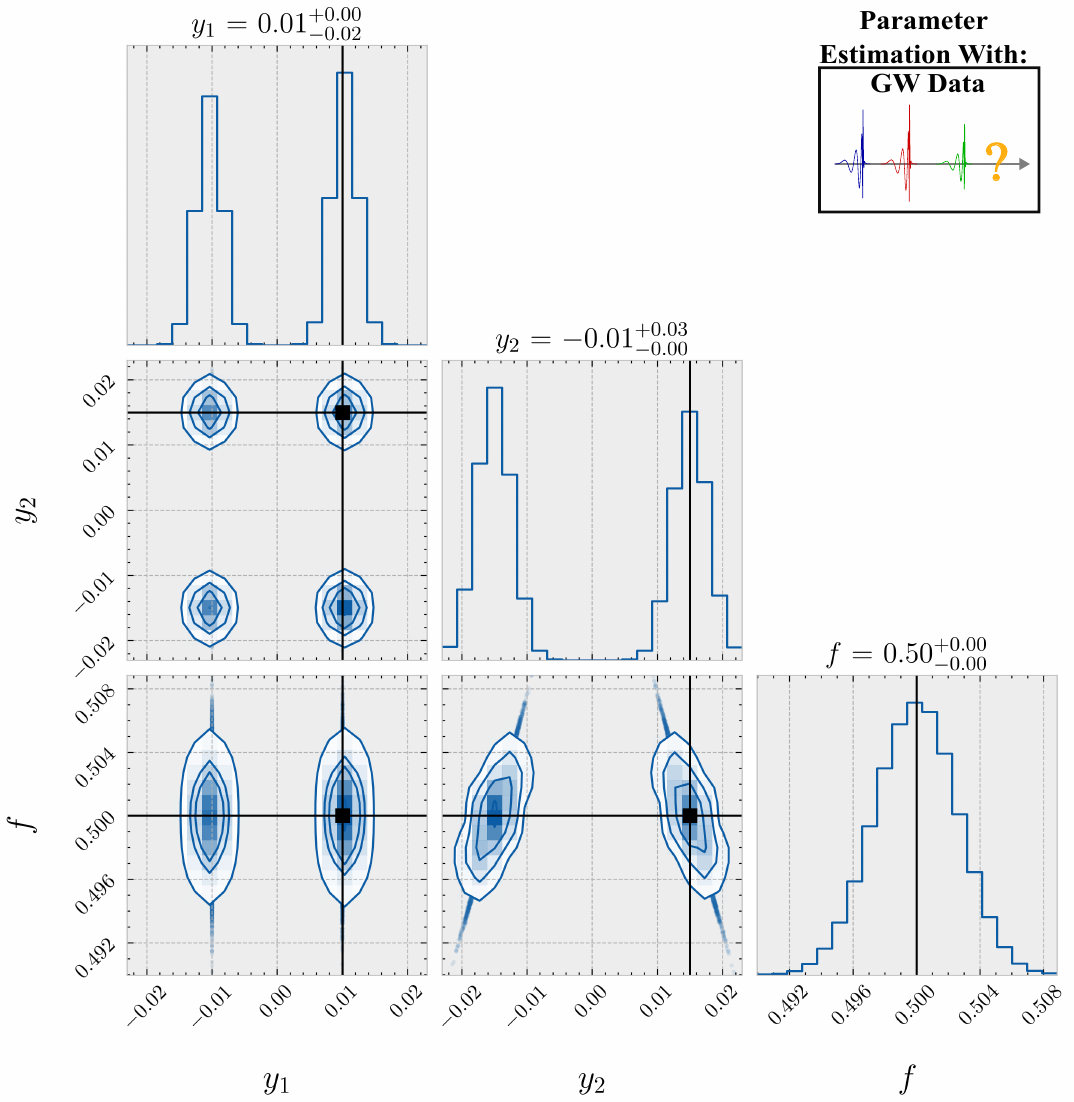}
\caption{Result of SIE lens reconstruction of dark siren lensed GW using injected parameters shown in Table \ref{table:dark_siren_lens_reconstruction_injection_setup} and the observables of the first three images from Table \ref{table:dark_siren_lens_reconstruction_injected_observables}, which an uncorrelated Gaussian likelihood with $0.01\%$ error on observed relative time delays and effective luminosity distances is assumed. After taking similarity transformation degeneracy into account, the five free parameters in the inference are $\{\vec{y}=(y_1, y_2), f, T_*, d_L\}$, which are the dimensionless source position, the SIE axis ratio, the time delay scaling $T_*\equiv\frac{1+z_l}{c}\frac{D_l D_s}{D_{ls}} \theta_E^2$ and the (unlensed) luminosity distance respectively. In this figure, we show the posterior distribution of the parameters that represent the dimensionless lens system: $\vec{y}=(y_1, y_2)$ and $f$. We can see, the parameters are recovered fairly well. We note that the dimensionless source position suffers from multiple solutions only because SIE lens has two degrees of reflectional symmetry. Other than that, there is no degeneracy among the parameters.
}
\label{STD 3 img source vs f (0.01percent error)}
\centering
\end{figure}

Fig. \ref{STD 3 img source vs f (5percent error)} and Fig. \ref{STD 3 img source vs f (0.01percent error)} show the result of the inference of the lens system's dimensionless parameters using the first three images, with the difference being they assume a $5\%$ and $0.01\%$ error on the (uncorrelated) Gaussian likelihood of the observables respectively. The full results are also shown in Fig. \ref{STD 3 img standard PE (5percent error)} and Fig. \ref{STD 3 img standard PE (0.01percent error)} in the Appendix respectively. From Fig. \ref{STD 3 img source vs f (5percent error)}, we can see the injection overlaps with the posterior, although the uncertainties are quite broad. From Fig. \ref{STD 3 img source vs f (0.01percent error)}, which the error of the Gaussian likelihood is small and the parameter estimation uncertainty is minimized, we recover the injection value accurately. This result is also expected, as the five degrees of information in the observables ($d_{L,1}^\text{eff}, d_{L,2}^\text{eff}, d_{L,3}^\text{eff}, \Delta t_{d,12}$ and $\Delta t_{d,23}$) match the five degrees of freedom in the irreducible free parameters ($\{\vec{y}=(y_1, y_2), f, T_*, d_L\}$). However, we note that the inference accuracy will not be as good as the case of using all four images. Nonetheless, it may still allow us to predict the image property of the fourth image and help for preparing or searching for the fourth image in the GW data, given that we identify the order of the images correctly, as discussed in \citet{ng2024uncovering}.

\subsection{Science Application: Hubble Constant Measurement}\label{S4.4}
Here, we review the procedures of measuring Hubble constant from lensed GW in the context of the various degeneracies. Demonstrations of such Hubble constant measurements have been done by various authors in the context of supplementary EM data \citet{Liao_2017, Hannuksela_2020} and statistical studies \citet{Jana_2023}. 

Suppose one has access to the EM counterpart or observation of the lens galaxy and the lensed host galaxy of the GW source, for example, through an archival search or a rapid EM follow-up~\citet{Hannuksela_2020,smith2023discovering}. In this case, we have access to the source and lens redshifts ($z_s$ and $z_l$). With this additional information, one can measure the Hubble constant $H_0$ as follows. Since the (unlensed) luminosity distance of the source $d_L$ and angular diameter distance to the source $D_s$ are related by $D_s={d_L}/{(1+z_s)^2}$, $T_*$ can be written as
\begin{equation}
    T_*=\frac{1+z_l}{c}\frac{D_l}{D_{ls}}\frac{d_L}{(1+z_s)^2} \theta_E^2=d_L \left[\frac{D_l}{D_{ls}}\frac{1+z_l}{c(1+z_s)^2} \theta_E^2\right].
\end{equation}
For our flat universe, we note that the Hubble constant information is encoded in the luminosity distance $d_L$ as shown in Eq.(\ref{d_L eq}). Then, we write
\begin{equation}
\begin{split}
    \Delta t_{d,ij} &=T_* \tau_{ij}= d_L \left[\frac{D_l}{D_{ls}}\frac{1+z_l}{c(1+z_s)^2} \theta_E^2\right] \tau_{ij}
    \\&= d_L G \tau_{ij},
\end{split} 
\end{equation}
where $G$ is defined as 
\begin{equation}
    G=G(z_l, z_s, \theta_E)\equiv\frac{D_l}{D_{ls}}\frac{1+z_l}{c(1+z_s)^2} \theta_E^2.
    \label{definiton of G}
\end{equation}
We stress that $G$ is independent of the Hubble constant (as it cancels out in the factor ${D_l}/{D_{ls}}$). In addition, some may argue that the angular Einstein radius $\theta_E$ may be dependent on $H_0$, for instance, for a point mass lens with mass $M$, $\theta_E=\sqrt{{4GMD_{ls}}/{c^2 D_l D_s}}$. But we note that for some lenses, such as the SIE (and SIS) that we are considering here, $\theta_E={4 \pi v^2 D_{ls}}/{c^2 D_s}$ is independent of the Hubble constant. Therefore, one should treat $\theta_E$ as just an angular scale of the lens that one can observe with an EM telescope \citep{Gonz_lez_2020}.

To summarize, from lensed GWs, we would obtain $\Delta t_{d,ij} = d_L G \tau_{ij}$ and $d_{L,i}^\text{eff}={d_L}/{\sqrt{|\mu_i|}}$. Previously, we mentioned we would be inferring $T_*$ and $d_L$ in GW lens reconstruction, but here, we note it being equivalent to inferring the value of $d_L$ and $G$ instead. This would collectively describe the degeneracies outlined in Section \ref{S4.1}, especially the partial degeneracy as a result of $\{z_l, z_s, \theta_E\}$ being coupled in the two separate quantities $d_L(z_s, H_0)$ and $G(z_l, z_s, \theta_E)$.
 
If the gravitational lens system is identified in the EM band, the EM telescope image data, as well as source and lens redshift information from photometric/spectroscopic redshift measurement, will yield a measurement for $z_s$, $z_l$ and $\theta_E$ directly. In principle, it may be possible, in certain rare scenarios, that the lens system is identified without identification of the source galaxy. In such a case, $z_l$ and $\theta_E$ might be known and one would be able to solve for $z_s$ from the inferred value of $G$. That is, $z_s$ can be inferred and is not required for cosmological applications. Indeed, one can solve for $H_0$ from the inferred luminosity distance $d_L(z_s, H_0)$ together with the time-delay distance. That is, the source redshift is not strictly speaking required for cosmological applications. 

\begin{figure}
\centering
\includegraphics[scale=0.45]{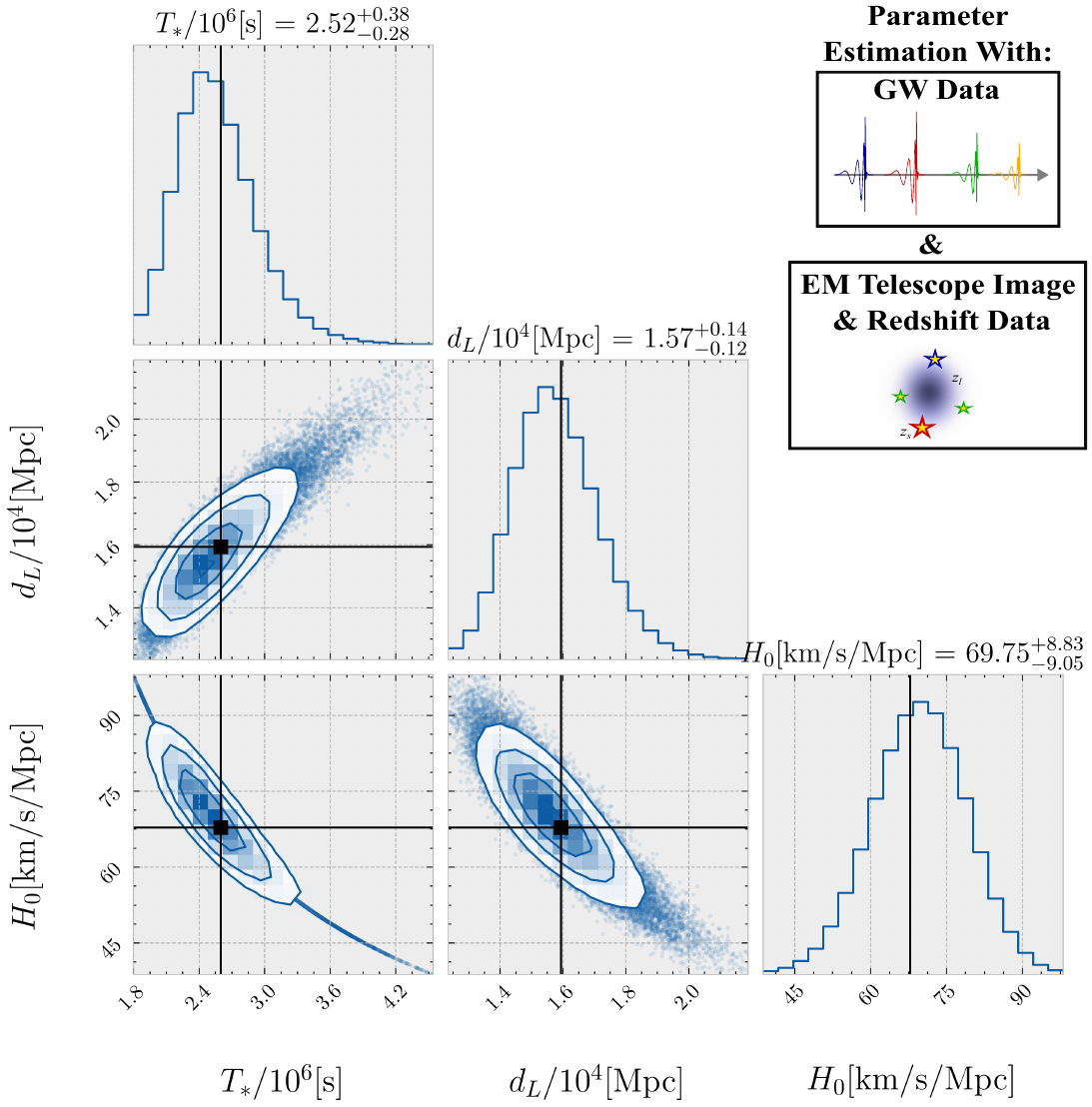}
\caption{Result of Hubble constant $H_0$ inference using result from the run shown in Fig. \ref{fig:STD_source_vs_f}, Fig. \ref{fig:STD_T_vs_dL} and Fig. \ref{fig:STD_standard_PE}, after assuming accurate information of $z_l=0.5, \theta_E\approx 0.592 \text{ arcseconds}$ and $z_s=2$ obtained from EM observation and lens system identification. The parameters shown: $\{T_* \text{ (in seconds)}, d_L \text{ (in Mpc)}, H_0 \text{ (in km} \text{ s}^{-1} \text{ Mpc}^{-1}\text{)}\}$ are the time delay scaling, the luminosity distance and Hubble constant respectively. We note that the peak of the inferred $H_0$ overlaps with the injection value of $67.8 \text{ km} \text{ s}^{-1} \text{ Mpc}^{-1}$.}
\label{fig:STD_T_vs_dL vs H0}
\centering
\end{figure}

To illustrate this, we perform a post-lens reconstruction analysis using the result from the last section (\textit{i.e.} the posterior $P(\vec{y}, f, T_*, d_L|\vec{d}_{GW})$ showed in Fig. \ref{fig:STD_source_vs_f}, Fig. \ref{fig:STD_T_vs_dL} and Fig. \ref{fig:STD_standard_PE}), from which we infer $H_0$. We assume the correct lens system is identified from EM observations, and we further assume the precise value of $z_l=0.5, \theta_E\approx 0.592 \text{ arcseconds}$ and $z_s=2$ are obtained without uncertainty. Fig. \ref{fig:STD_T_vs_dL vs H0} shows the result, in which we recovered the injected value of the Hubble constant without source redshift. Therefore, the similarity transformation degeneracy can be broken, and the Hubble constant can be measured when one can observe the lens and the host galaxy through another channel.

In our discussion, we stated that, due to the similarity transformation degeneracy, we could only perform precise cosmological studies in GW lensing when complemented with EM information (source and lens redshift, as well as Einstein radius). But we stress that, it is still possible to constrain cosmological parameters (such as $H_0$) to some degree if a known statistical distribution of the source and lens redshift ($z_s$ and $z_l$), as well as the velocity dispersion $v$ of the lens are applied in the analysis (such statistical distribution can be obtained by non-lensing related survey). Such methodology is demonstrated in \citet{Jana_2023} (in which they also studied the constraint on the matter density $\Omega_m$). In our work, we restrict ourselves to the case of precise measurements, where we do not use the statistical distributions of the source and lens parameters $z_s$, $z_l$ and $v$. This explains the apparent discrepancy between our conclusions. 

\subsection{Analytical Example: SIS Lens}\label{S4.5}
Here, we provide an analytical example to explicitly demonstrate the idea of dark siren GW lens reconstruction under similarity transformation degeneracy. Consider a singular isothermal sphere (SIS) lens, with its surface density given by 
\begin{equation}
    \Sigma(\vec{\xi})=\frac{v^2}{2\xi},
\end{equation}
where $v$ is the velocity dispersion of the lens galaxy and $\vec{\xi}$ is the (Cartesian) impact parameter on the lens plane (\textit{i.e.} $\vec{\xi}=\vec{x} \theta_E D_l$ in our notation). Its (angular) Einstein radius is given by $\theta_E = 4\pi v^2 {D_{ls}}/{D_s}$ and the dimensionless deflection potential is $\Psi(\vec{x})=\lvert\vec{x}\rvert$ \citep{Takahashi_2003}.

Therefore, the dimensionless lens equation is now
\begin{equation}
    \vec{y}=\vec{x} - \nabla_{\vec{x}}\lvert\vec{x}\rvert=\vec{x} \left( 1 - \frac{1}{\lvert\vec{x}\rvert}\right).
\end{equation}

Since this is a symmetric system, we expect the source and images to lie on the same line (which goes through the centre of the lens). Hence it becomes a one dimensional problem: $\vec{y}\rightarrow y$, $\vec{x}\rightarrow x$. Moreover, we can set $y>0$ without loss of generality. Then, the dimensionless lens equation becomes
\begin{equation}
    y = x \left( 1 - \frac{1}{\lvert x\rvert}\right),
\end{equation}
which has two solutions for the dimensionless image position when $0<y<1$:
\begin{equation}
\begin{split}
    x_+ &= y + 1, \\
    x_- &= y - 1.
\end{split}
\end{equation}

We then have the time delay of the $x_+$ image as:
\begin{equation}
\begin{split}
t_d(x_+) &= \frac{1+z_l}{c}\frac{D_l D_s}{D_{ls}} \theta_E^2\left[\frac{|y + 1-y|^2}{2}-(y + 1)\right] \\
&=\frac{1+z_l}{c}\frac{D_l D_s}{D_{ls}} \theta_E^2 \left( -y-\frac{1}{2}\right),
\end{split}
\end{equation}

and that of the $x_-$ image as:
\begin{equation}
    t_d(x_-) = \frac{1+z_l}{c}\frac{D_l D_s}{D_{ls}} \theta_E^2 \left( y-\frac{1}{2}\right).
\end{equation}
Hence, the relative time delay between the two images is
\begin{equation}
    \Delta t_d = \frac{1+z_l}{c}\frac{D_l D_s}{D_{ls}} \theta_E^2 (2y) = T_* (2y).
    \label{SIS td eq}
\end{equation}
Meanwhile, the image magnifications are given by 
\begin{equation}
    \mu_\pm = \pm 1 + {1}/{y}.
\end{equation}
The corresponding effective luminosity distances are
\begin{equation}
    d_{L,\pm}^\text{eff} = {d_L}/{\sqrt{\lvert \mu_\pm \rvert}}={d_L}/{\sqrt{\lvert \pm 1 + {1}/{y} \rvert}}.
\end{equation}

Now, assume we only have dark siren lensed GW observation so that we only observe $\Delta t_d$ and $d_{L,\pm}^\text{eff}$. Furthermore, assume the observables we obtained are certain. One can then solve for the dimensionless source position $y$ fully in terms of the observables: by realising ${\mu_+}/{\mu_-}=-\left({d_{L,-}^\text{eff}}/{d_{L,+}^\text{eff}}\right)^2$, we have 
\begin{equation}
    y=\frac{\mu_+/\mu_--1}{\mu_+/\mu_-+1} = \frac{\left({d_{L,-}^\text{eff}}/{d_{L,+}^\text{eff}}\right)^2+1}{\left({d_{L,-}^\text{eff}}/{d_{L,+}^\text{eff}}\right)^2-1}.
\end{equation}
Then, from Eq.(\ref{SIS td eq}), $T_*$ can be calculated from the observables directly as
\begin{equation}
    T_*=\frac{\Delta t_d}{2y}=\frac{\Delta t_d}{2}\frac{\left({d_{L,-}^\text{eff}}/{d_{L,+}^\text{eff}}\right)^2-1}{\left({d_{L,-}^\text{eff}}/{d_{L,+}^\text{eff}}\right)^2+1},
\end{equation}
and $d_L$ can be calculated by
\begin{equation}
    d_L=d_{L,\pm}^\text{eff}\sqrt{\left| \pm 1 + \frac{\left({d_{L,-}^\text{eff}}/{d_{L,+}^\text{eff}}\right)^2-1}{\left({d_{L,-}^\text{eff}}/{d_{L,+}^\text{eff}}\right)^2+1} \right|}.
\end{equation}
At this stage, we have used up all the information we have, finding the value of $T_*$ but not being able to find the values of its components, including $\theta_E$ (which is then related to the velocity dispersion $v$ of the galaxy). Therefore, in a more general dark siren SIS lens reconstruction scenario, we would only be able to infer $P(y, T_*, d_L|\vec{d}_{GW})$, instead of the full joint posterior distribution $P(v,z_l, \beta, z_s, H_0|\vec{d}_{GW})$, which would be the optimal result.

\section{Mass-sheet degeneracy} \label{S5}
The mass-sheet degeneracy (or magnification degeneracy) was first pointed out by \citet{1985ApJ...289L...1F}. It refers to a transformation of the lens system that leaves observables such as image positions and flux/magnification ratio of images invariant, except for the quantity $H_0 t_d$. This degeneracy is a well-known issue in lensing observation and analysis on the EM side, and it can lead to a biased estimation of the Hubble constant $H_0$ from lensing if unaccounted for \citep{Schneider_2013}.   

The mass sheet transformation modifies the convergence of lens model as follows \citep{Schneider_2013}:
\begin{equation}
    \bar{\kappa}(\vec{\theta}) = (1-\kappa_0) \kappa(\vec{\theta}) + \kappa_0,
\end{equation}
where $\bar{\kappa}$ denotes the transformed convergence of the new lens model and $\kappa$ denotes the original convergence of the original lens model. Physically, it can be interpreted as adding a uniform and infinitely-spanning mass sheet with convergence $\kappa_0$ at the lens while reducing the mass density of the original lens by the factor of $(1-\kappa_0)$. The presence of such mass sheet may be due to external convergence, such as the lens being embedded in low-mass galaxy cluster, or due to structures along the line of sight \citep{1985ApJ...289L...1F, Schneider_2013, Birrer_2016}. \citet{Schneider_2013} also demonstrated that the infinitely spanned uniform mass sheet has very similar effect as a finite mass sheet that is located within $2\theta_E$ from the centre of the lens, hence the mass sheet transformation can also approximate a similar transformation of the lens model just near the core region of the lens. Note that the Einstein radius $\theta_E$ is not modified \footnote{In fact, for a circular lens, the mass-sheet degeneracy also preserves the total mass within the Einstein radius \citep{Saha_2000}.}. We also note that, the transformed dimensionless deflection potential, $\bar{\Psi}(\vec{x})$, becomes:
\begin{equation}
    \bar{\Psi}(\vec{x})=(1-\kappa_0) \Psi(\vec{x}) + \frac{\kappa_0}{2} |\vec{x}|^2.
\end{equation}

The transformation also changes the old source position $\vec{\beta}$ to the new source position $\vec{\bar{\beta}}$ by
\begin{equation}
    \vec{\bar{\beta}} = (1-\kappa_0) \vec{\beta}.
\end{equation}
In dimensionless coordinates, the transformation is simply 
\begin{equation}
    \vec{\bar{y}} = (1-\kappa_0)\vec{y},
\end{equation}
where $\vec{\bar{y}}$ is the new dimensionless source position.

Therefore, the transformed lens equation is:
\begin{equation}
    \vec{\bar{y}}=\vec{x}-\nabla_{\vec{x}} \bar{\Psi}(\vec{x}).
\end{equation}
 Directly substituting the transformation above, we have the following steps:
\begin{align*} 
     (1-\kappa_0)\vec{y} &= \vec{x}-\nabla_{\vec{x}} \left[(1-\kappa_0) \Psi(\vec{x}) + \frac{\kappa_0}{2} |\vec{x}|^2\right] \\
     &= \vec{x}-(1-\kappa_0) \nabla_{\vec{x}} \Psi(\vec{x}) - \frac{\kappa_0}{2} \nabla_{\vec{x}} |\vec{x}|^2 \\
     &= \vec{x}-(1-\kappa_0) \nabla_{\vec{x}} \Psi(\vec{x}) - \kappa_0 \vec{x} \\
     &= (1-\kappa_0) \vec{x}-(1-\kappa_0) \nabla_{\vec{x}} \Psi(\vec{x}),
\end{align*}
which, after cancelling $(1-\kappa_0)$ from both sides, reduces to the original lens equation of the system before mass sheet transformation: $\vec{y} = \vec{x}-\nabla_{\vec{x}} \Psi(\vec{x})$. Hence, the solution of dimensionless image position $\vec{x}$ remains the same, and since the Einstein radius is not changed, the actual image positions of the lens system are not altered by the mass sheet transformation. Therefore, it is impossible to lift the degeneracy from image position alone, even in the case of EM lensing observation \citep{alma991039610282103407}.

Under the mass sheet transformation, the new image magnification is 
\begin{equation}
    \bar{\mu}_i =\frac{\mu_i}{(1-\kappa_0)^2},
\end{equation}
where $\mu$ is the original magnification before mass sheet transformation and the subscript $i$ denotes the $i$-th image. We note that, since all image magnifications are affected by the same factor of $(1-\kappa_0)^{-2}$, the magnification ratios are preserved under mass sheet transformation
\begin{equation}
    \frac{\bar{\mu}_i}{\bar{\mu}_j} = \frac{\mu_i}{\mu_j}.
\end{equation}
Hence, in the case of EM lensing observation, as long as the actual source flux is unknown and we only have access to flux ratio between images, we cannot lift the mass-sheet degeneracy.

For GW lensing, since it is a standard siren, in the case of no lens and source redshift measurement, after mass sheet transformation, we will be measuring effective luminosity distance of the images:
\begin{equation}
    \bar{d}^\text{eff}_{L,i}=\frac{d_{L}}{\sqrt{|\bar{\mu}_i|}}=\frac{d_{L}(1-\kappa_0)}{\sqrt{|\mu_i|}}=\frac{d_L^{\kappa_0}}{\sqrt{|\mu_i|}}
    \label{dL^eff after MST definition}
\end{equation}
where we define $d_L^{\kappa_0}\equiv(1-\kappa_0) d_L$ as the mass-sheet-degenerated luminosity distance. We also note that the Morse index of the images remain unchanged.

\begin{figure*}
\centering
\includegraphics[scale=0.42]{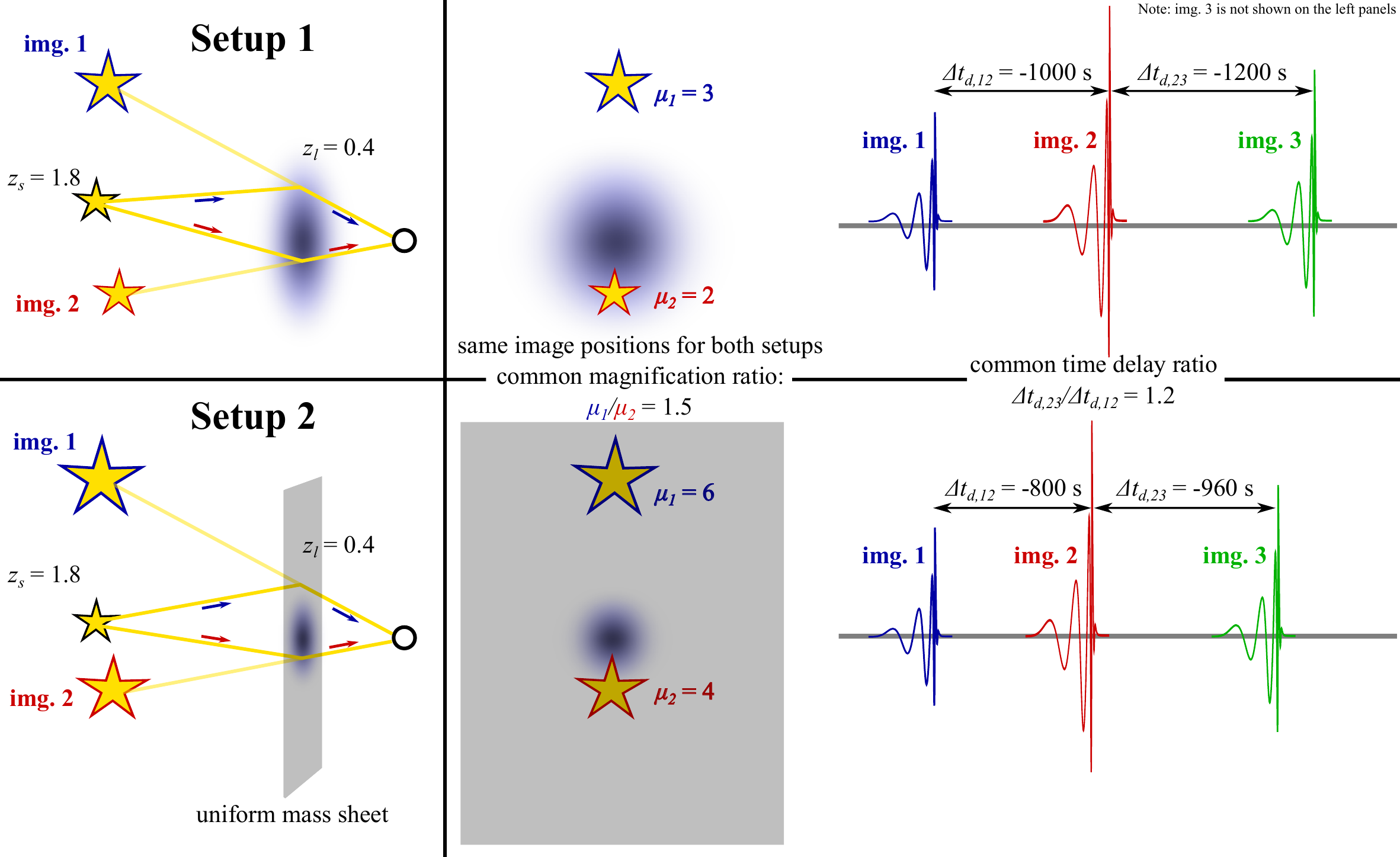}
\caption{Illustrative example of mass-sheet degeneracy in multi-messenger lensing: the upper and lower lens system on the left panel corresponds to a lens system before and after mass sheet transformation respectively. We note that the source emits both EM waves and GW waves (for instance, a binary neutron star merger, or a binary black hole merger and its host galaxy). The middle panel illustrates the image position and magnification observed through EM telescope, and the right panel illustrates the relative time delay between images observed by GW detectors. Due to mass-sheet degeneracy, the two systems lead to the same image positions, flux ratio and relative time delay ratio of different image pairs.}
\label{mass-sheet degeneracy illustration}
\centering
\end{figure*}

Moreover, the new relative time delay is
\begin{equation}
    \Delta\bar{t}_{d,ij}=(1-\kappa_0)\Delta t_{d,ij},
\end{equation}
where $\Delta t_{d,ij}=T_* \tau_{ij}$ is the original relative time delay between image $i$ and $j$, and again, we emphasize that $\tau_{ij}$ is difference of the dimensionless time delay for image pair $i,j$ of the original system. Hence, we can define the mass-sheet-degenerated time delay scaling as
\begin{equation}
    T_*^{\kappa_0}\equiv(1-\kappa_0)T_*
\end{equation}
We can then write the new relative time delay as
\begin{equation}
    \Delta\bar{t}_{d,ij}=T_*^{\kappa_0}\tau_{ij}.
\end{equation}
In addition, we note that, similar to the invariant magnification ratio of different images, the relative time delay ratios of different image pairs:
\begin{equation}
    \frac{\Delta\bar{t}_{d,ij}}{\Delta\bar{t}_{d,kl}}=\frac{\tau_{ij}}{\tau_{kl}},
\end{equation}
are also invariant under mass sheet transformation due to the common factor of $T_*^{\kappa_0}$ on all time delays. Fig. \ref{mass-sheet degeneracy illustration} summarizes the physical picture and the invariant observables of a multi-messenger lens system under mass sheet transformation.

\subsection{Lens Reconstruction with Similarity Transformation and mass-sheet degeneracy}
Here, for dark siren GW lensing, we can only measure relative time delays $\Delta\bar{t}_{d,ij}$ and effective luminosity distances $\bar{d}^\text{eff}_{L,i}$ of the images. In lens reconstruction, since $T_*^{\kappa_0}$ and $d_L^{\kappa_0}$ are common factor of all the images' relative time delay and effective luminosity distance respectively, similar to the case of similarity transformation degeneracy, we can eliminate them and obtain the magnification ratios as
\begin{equation}
    \frac{\mu_i}{\mu_j}=\left(\frac{\bar{d}_{L,j}^\text{eff}}{\bar{d}_{L,i}^\text{eff}}\right)^2 (-1)^{\Delta n_{ij}},
\end{equation}
and the dimensionless time delay ratios as
\begin{equation}
     \frac{\tau_{ij}}{\tau_{mn}} = \frac{\Delta \bar{t}_{d,ij}}{\Delta \bar{t}_{d,mn}}.
\end{equation}
From these ratios, we can only reconstruct the dimensionless system that does not have mass sheet transformation taken into account, namely the original (pre-mass-sheet-transformation) lens profile $\Psi(\vec{x})$  (for instance, the axis ratio $f$ in SIE lens) and also the original dimensionless source position $\vec{y}$. Then, we can only infer the values of the mass-sheet-degenerated time delay scaling $T_*^{\kappa_0}$ and the mass-sheet-degenerated luminosity distance $d_L^{\kappa_0}$, and we emphasize that we cannot resolve any terms inside of them. In summary, we can only reconstruct $P(\vec{y}, \Psi(\vec{x}), T_*^{\kappa_0}, d_L^{\kappa_0}|\vec{d}_{GW})$ with dark siren GW lensing.

\begin{table*}
\centering
\begin{tabular}{|c | c | c | c|} 
 \hline
 Type of parameters & $\vec{\Phi}_l$ (SIE lens with mass sheet) & $\vec{\Phi}_s$ (GW source) & $\vec{\Phi}_{c}$ (cosmology) \\ [0.5ex] 
 \hline
  Injection values & $z_l=0.5, v=0.0006c, f = 0.5, \kappa_0=0.05$ & $z_s=2, \vec{\beta}\approx(0.00592,0.00887) \text{ (in arcseconds)}$ & $H_0=67.8 \text{ km} \text{ s}^{-1} \text{ Mpc}^{-1}$ \\
 \hline
\end{tabular}
\caption{Injection values of the system parameters for our lens reconstruction example under mass sheet transformation: 1. for the lens parameters, $z_l, v, f, \kappa_0$ are the redshift, velocity dispersion, SIE axis ratio and mass sheet convergence respectively. Note that $c$ is the speed of light and we align the long axis of the SIE with the vertical axis (\textit{i.e.} the $y_2$ or $x_2$ axis). 2. For the source, $z_s, \vec{\beta}$ are the redshift and the (angular) source position in arcseconds (before mass sheet transformation is applied) respectively. 3. For cosmology part, $H_0$ is the Hubble constant.}
\label{table: dark siren MST lens reconstruction injection setup}
\end{table*}

\begin{table*}
\centering
\begin{tabular}{|c | c | c|} 
 \hline
 Observables & Effective luminosity distance: $d_{L,1}^\text{eff}, d_{L,2}^\text{eff}, d_{L,3}^\text{eff}, d_{L,4}^\text{eff}$ (in Mpc) & Relative time delay between images: $\Delta t_{d,12}, \Delta t_{d,23}, \Delta t_{d,34}$ (in s) \\ [0.5ex] 
 \hline
 Injected values & $8934,8775,11966,12510$ & $-53026,-467148,-63264$\\
 \hline
\end{tabular}
\caption{GW lensing observables of injected system setup shown in Table \ref{table: dark siren MST lens reconstruction injection setup}. Note that we have four images and the image types are I-I-II-II.}
\label{table: dark siren MST lens reconstruction injected observables}
\end{table*}

\begin{figure}
\centering
\includegraphics[scale=0.62]{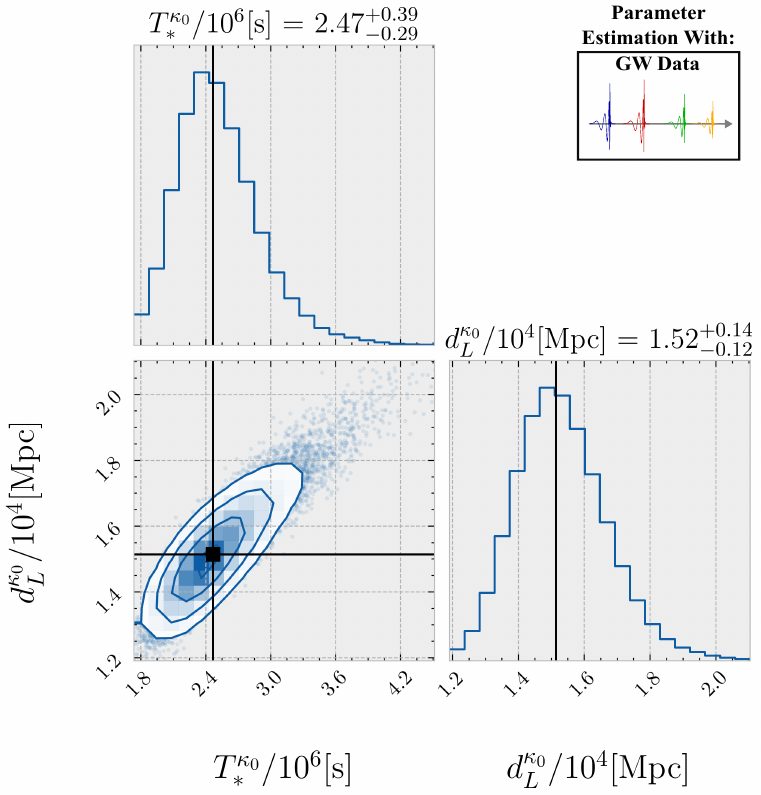}
\caption{Result of SIE lens reconstruction of dark siren lensed GW using injected parameters shown in Table \ref{table: dark siren MST lens reconstruction injection setup} and Table \ref{table: dark siren MST lens reconstruction injected observables}. After taking similarity transformation degeneracy and mass-sheet degeneracy into account, the five free parameters are $\{\vec{y}=(y_1, y_2), f, T_*^{\kappa_0}, d_L^{\kappa_0}\}$, which are the dimensionless source position (before mass sheet transformation), the SIE axis ratio, the mass-sheet-degenerated time delay scaling $T_*^{\kappa_0}\equiv (1-\kappa_0) \frac{1+z_l}{c}\frac{D_l D_s}{D_{ls}} \theta_E^2$ and the mass-sheet-degenerated luminosity distance respectively. Here, we show the posterior distribution for the parameters that describe the scale of the system: $T_*^{\kappa_0} \text{ (in seconds)}, d_L^{\kappa_0} \text{ (in Mpc)}$. We can see, the two parameters are recovered fairly well without degeneracy.}
\label{MSD T vs dL}
\centering
\end{figure}

Here we show an example of dark siren GW lens reconstruction under both similarity transformation degeneracy and mass-sheet degeneracy. Similar to Section \ref{S4}, we inject an SIE lens with the system parameters shown in Table \ref{table: dark siren MST lens reconstruction injection setup}, which gives injected observables presented in Table \ref{table: dark siren MST lens reconstruction injected observables}. Same as the likelihood $P(\vec{d}_{GW}|\vec{\nu}_{im})$ described in Section \ref{S4.2}, we assume an uncorrelated 5\% Gaussian error on the measured $d_{L,j}^\text{eff}$ and $\Delta t_{d,ij}$. Fig. \ref{MSD T vs dL} shows the result of lens reconstruction of the free parameters $\{T_*^{\kappa_0}, d_L^{\kappa_0}\}$ (the full result of parameter estimation of the parameters $\{\vec{y}, f, T_*^{\kappa_0}, d_L^{\kappa_0}\}$ is shown in Fig. \ref{MSD standard PE} in the Appendix), and we are able to recover them accurately. However, it is not possible to distinguish the effect of mass sheet convergence $\kappa_0$ from the absolute time-delay scaling $T_*$ or luminosity distance $d_L$. We note that, in Section III A of \citet{Cremonese_2021}, they also demonstrated the scaling relation of mass sheet density and the GW amplitude (which is related to the effective luminosity distance defined in Eq. (\ref{dL^eff after MST definition})), as well as mass sheet degeneracy itself from the perspective of GW waveform in the geometric optics.

\subsection{Inability to Resolve mass-sheet degeneracy Even with a Telescope Image and Spectroscopic Redshifts}\label{S5.2}
Similar to Section \ref{S4.4}, we assume we obtain the source and lens redshift ($z_s$ and $z_l$) through identifying the corresponding lens system in EM observations. We furthermore assume that the Einstein radius $\theta_E$ can be obtained from a high-resolution image of the lens. Since GW is a standard siren, it may intuitively make sense that one should be able to break the mass-sheet degeneracy and measure both $\kappa_0$ and $H_0$ at the same time. Unfortunately, this is not the case. Under the mass sheet transformation, the relative time delay between images are given by
\begin{equation}
\begin{split}
    \Delta\bar{t}_{d,ij} &=(1-\kappa_0)T_* \tau_{ij}=\left[(1-\kappa_0) d_L \right] \left[\frac{D_l}{D_{ls}}\frac{1+z_l}{c(1+z_s)^2} \theta_E^2 \right] \tau_{ij}
    \\&= d_L^{\kappa_0} G \tau_{ij},
\end{split}
\label{mass sheet degenerated td equation}
\end{equation}
where $d_L^{\kappa_0}=d_L^{\kappa_0}(z_s, H_0, \kappa_0)\equiv(1-\kappa_0) d_L(z_s, H_0)$ and $G(z_l, z_s, \theta_E)$ is defined in Eq. (\ref{definiton of G}). In a flat universe, we have
\begin{equation}
    d_L^{\kappa_0}= \frac{c(1-\kappa_0)}{H_0} \int_0^{z_s}\frac{dz'}{\sqrt{\Omega_r(1+z')^4+\Omega_m(1+z')^3+\Omega_\Lambda}}.
    \label{d_L^kappa_0 equation}
\end{equation}

Thus, after mass sheet transformation, $\bar{d}^\text{eff}_{L,i}=\frac{d_L^{\kappa_0}}{\sqrt{|\mu_i|}}$ and $\Delta\bar{t}_{d,ij}= d_L^{\kappa_0} G \tau_{ij}$. Recall that for the absolute scale of the lens system, we can only infer the values of the pair $(T_*^{\kappa_0}, d_L^{\kappa_0})$. Given the relationship between the luminosity distance and the angular diameter distance to the source, we can conclude that, effectively or equivalently, one can only infer the values of the pair $d_L^{\kappa_0}(z_s, H_0, \kappa_0)$ and $G(z_l, z_s, \theta_E)$.

As mentioned in Section \ref{S4.4}, with EM information, we will have $z_s$, $z_l$ and $\theta_E$ information. Under fixed cosmology, we can infer $G(z_l, z_s, \theta_E)$. However, after we infer the value of $d_L^{\kappa_0}(z_s, H_0, \kappa_0)$, even with knowledge of $z_s$, we still have a coupled pair of unknown $(H_0, \kappa_0)$ hidden in $d_L^{\kappa_0}$ in the form of ${(1-\kappa_0)}/{H_0}$ (Eq. (\ref{d_L^kappa_0 equation})), making the two fully degenerate. The reason behind is the effect of mass sheet transformation modifies the effective luminosity distance and time delay in the same way, namely multiplying the factor $(1-\kappa_0)$. Therefore, lensed GWs with EM counterpart (or observation of the lens and the lensed host galaxy) will not be able to break mass-sheet degeneracy with just lensing observables. 

\begin{figure}
\centering
\includegraphics[scale=0.62]{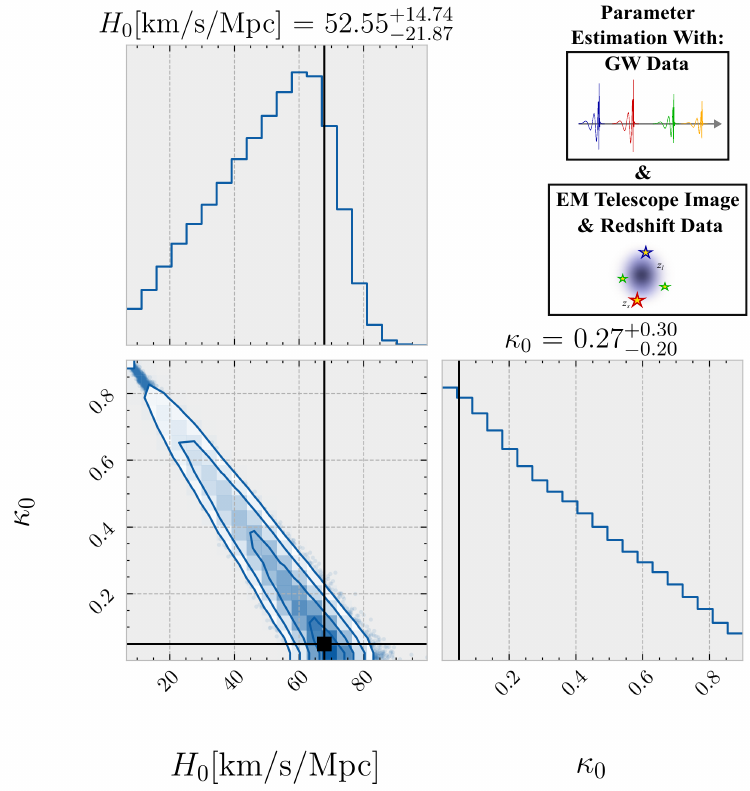}
\includegraphics[scale=0.62]{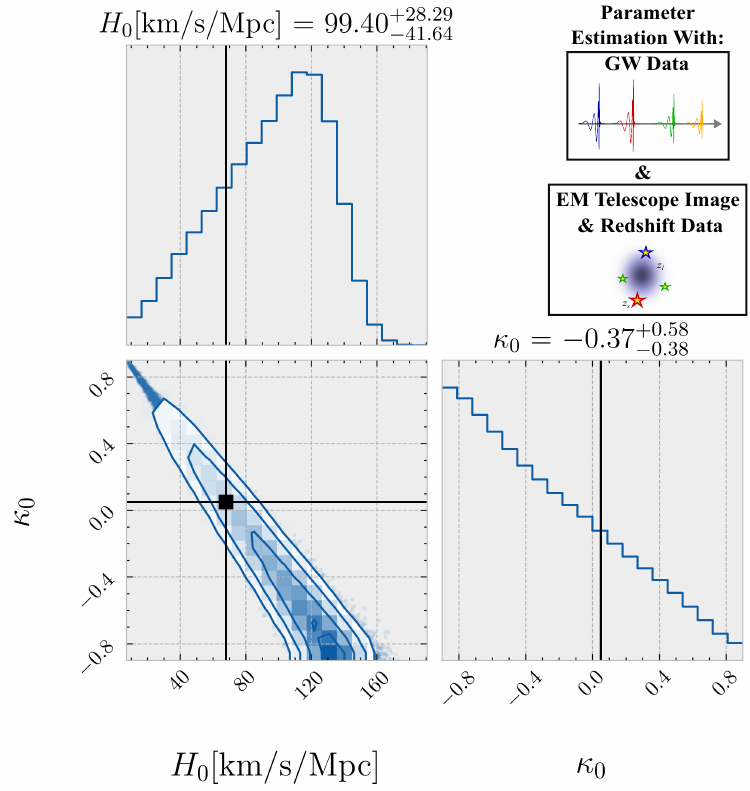}
\caption{Result of the Hubble constant $H_0$ and mass sheet convergence $\kappa_0$ inference for lensed GWs. Injection model and parameters shown in Table \ref{table: dark siren MST lens reconstruction injection setup} and Table \ref{table: dark siren MST lens reconstruction injected observables} are used. In the run presented in the upper panel, the prior of $\kappa_0$ is set to be uniform from $0$ to $0.9$, while that of the lower panel is set to be uniform from $-0.9$ to $0.9$. We assume $z_s$, $z_l$ and $\theta_E$ are known from identifying the corresponding lens system in EM telescope observation. As we can see, the Hubble constant $H_0$ and mass sheet convergence $\kappa_0$ are poorly constrained, degenerate, and sensitive to the choice of prior.}
\label{MST degeneracy case H0 vs kapp0 with kappa0 different prior}
\centering
\end{figure}

Fig. \ref{MST degeneracy case H0 vs kapp0 with kappa0 different prior} illustrates parameter estimation attempts trying to infer $\kappa_0$ and $H_0$ with a known (fixed) $z_s$, $z_l$ and $\theta_E$. Both panels have the same setup as the one described in Tables \ref{table: dark siren MST lens reconstruction injection setup} and \ref{table: dark siren MST lens reconstruction injected observables}. We assume the same uniform prior for $H_0$. The difference between the two panels is that the prior of $\kappa_0$ in the simulation shown in the upper panel is set to be uniform from $0$ to $0.9$, while the simulation shown in the lower panel is set to be uniform from $-0.9$ to $0.9$.\footnote{This is unphysical since we include the possibility of negative mass sheet, but mathematically, it is still valid, and we intend to use this prior to check the validity of the result shown in the upper panel.} For both results, we can see there is a $H_0$-$\kappa_0$ degeneracy spanning across the parameter space.

In the upper panel, it may look as if the peak of the posterior of $H_0$ still roughly overlaps with the injected values. However, it is just a coincidence due to assumption of priors for $H_0$ and $\kappa_0$.When we assume a different prior on $\kappa_0$, as shown in the lower panel, the posterior peak of $H_0$ is far off from injection, and the posterior probability of $\kappa_0$ drops as $\kappa_0$ increases, even in region where $\kappa_0$ is negative. It becomes clear that the inferred posterior distribution of $\kappa_0$ is not informative and affected by prior choice. These results indicate that constraining both $\kappa_0$ and $H_0$ simultaneously from GWs strongly lensed by galaxy is not possible even when an EM lens reconstruction is available, due to the mass-sheet degeneracy.

If we perform Hubble constant measurement ignoring the possible presence of a mass-sheet (as demonstrated in Section \ref{S4.4}), our measured Hubble constant would be biased from the true Hubble constant by a factor of $\frac{1}{1-\kappa_0}$. Such issues in Hubble constant measurement via observing lensing of EM waves are further discussed in \citet{Schneider_2013}.

In addition, the above analysis can be applied to similar situations in lensing of light. For instance, for time varying light sources that may not be standard candle, such as quasars, if one assume a fixed cosmology and known $H_0$, mass-sheet degeneracy can be broken and the mass sheet density $\kappa_0$ can be found as follows. Despite there will not be measurement of the effective luminosity distance $\bar{d}_L^{\text{eff}}$ (as the total flux is not known), from the image position and time delay measurement from the EM telescope, one should be able to measure the Einstein radius $\theta_E$, as well as reconstruct the dimensionless source position and calculate the dimensionless time delay $\tau_{ij}$ between images. Together with redshift measurement of the lens and source, $G(z_s, H_0, \theta_E)$ can be calculated, hence the value of $d_L^{\kappa_0}$ can be calculated with eq. (\ref{mass sheet degenerated td equation}). As cosmology is fixed and $z_s$ is known, $\kappa_0$ can be found from eq. (\ref{d_L^kappa_0 equation}) easily. This would be helpful if one would like to study the lens galaxy, such as it being possibly embedded in a galaxy cluster or a possible finite mass sheet located near the centre of the lens that can mimic the mass-sheet transformation \citep{1985ApJ...289L...1F, Schneider_2013, Birrer_2016}, without standard candle sources (but an accurate value of $H_0$ from other measurement is required). 

\section{Science applications and implications of the two degeneracies} \label{S6}
\subsection{Test of Modified Gravitational Wave Propagation}
Some theories modifying general relativity on cosmological scales have as typical signature a modified GW dispersion relation (propagation effect) \citep{Deffayet_2007, PhysRevD.98.023510, Belgacem_2019, PhysRevD.99.083504, Finke_2021}. Moreover, modified GW propagation may be one of the most promising window to observationally test such theories \citep{Mirshekari_2012, Finke_2021}. The idea of testing modified GW propagation effect on the luminosity distance of GW (which we will refer as GW luminosity distance, $d_L^\text{gw}$, below) by lensing was previously discussed in \citet{Finke_2021, iacovelli2022modified, narola2023modified}, on which the results presented in this section build. In addition, here we ask the question: "Given that the proposed modified GW propagation tests rely on luminosity distance measurements, could the mass-sheet degeneracy alter their results?" The answer appears to be no.

The relationship between GW and EM luminosity distance $d_L^\text{em}$ (which is just the original/standard luminosity distance as defined in Eq. (\ref{d_L eq})) is written as
\begin{equation}
    d_L^\text{gw}=\Xi(z_s) d_L^\text{em},
\end{equation}
where for a wide range of modified gravity theories, the GW to EM luminosity distance ratio $\Xi(z_s)$ can be parametrized as in \citet{Belgacem_2018}:
\begin{equation}
    \Xi(z_s)=\Xi_0 + \frac{1-\Xi_0}{(1+z_s)^n},
\end{equation}
where the parameters $\Xi_0$ and $n$ are determined by the specific theory \citep{Belgacem_2019}.

We now include mass-sheet degeneracy in our analysis, using same convention as Section \ref{S5}. The measured effective luminosity distance for the $i$-th GW lensed image is given by
\begin{equation}
    \bar{d}_{L,i}^\text{eff}=\frac{d_L^\text{gw}(1-\kappa_0)}{\sqrt{|\mu_i|}}=\frac{d_L^{\text{gw}, \kappa_0}}{\sqrt{|\mu_i|}},
    \label{MGW eff dL eq}
\end{equation}
where we defined the mass-sheet-degenerated GW luminosity distance as $d_L^{\text{gw}, \kappa_0} \equiv d_L^\text{gw}(1-\kappa_0)=\Xi(z_s) d_L^{em}(1-\kappa_0)$.

Meanwhile, the relative time delay between images is not affected, which means we still have $\Delta\bar{t}_{d,ij}=T_*^{\kappa_0}\tau_{ij}$ as before. Since the angular diameter distance to source and EM luminosity distance relation is $D_s={d_L^\text{em}}/{(1+z_s)^2}$, we have
\begin{equation}
    T_*^{\kappa_0} = \left[d_L^\text{em}(1-\kappa_0)\right] G(z_l, z_s, \theta_E).
    \label{MGW time delay scaling}
\end{equation}

Therefore, in dark siren GW lens reconstruction, besides from all the pre-mass sheet transformation dimensionless quantities of the lens system (\textit{i.e.} the dimensionless source position $\vec{y}$ and the SIE axis ratio $f$ in our example), we can only infer the values of $d_L^{\text{gw}, \kappa_0}$ and $T_*^{\kappa_0}$. Assume $z_l$, $z_s$ and $\theta_E$ are known from EM observation after identifying the lens system, we can calculate the value of $G(z_l, z_s, \theta_E)$ \footnote{We note that unlike in Section \ref{S4.4} or \ref{S5.2} where GW and EM luminosity distance are the same, $G(z_l, z_s, \theta_E)$ here cannot be inferred directly.}, and the GW to EM luminosity distance ratio can then be calculated by
\begin{equation}
    \Xi(z_s) = \frac{d_L^{\text{gw}, \kappa_0}}{T_*^{\kappa_0}}G(z_l, z_s, \theta_E),
    \label{Xi(z_s) calculation}
\end{equation}
which means that we do not even need to have knowledge on the Hubble constant $H_0$ or the mass sheet convergence $\kappa_0$. We only require a good lens reconstruction process that infer accurate values of $d_L^{\text{gw}, \kappa_0}$ and $T_*^{\kappa_0}$, as well as $z_l$, $z_s$ and $\theta_E$ information of the system. In other words, contrary to a naive expectation, this type of modified gravity test is immune to mass-sheet degeneracy. It can be understood as a result of the modified GW propagation only affecting the effective luminosity distance measurement (Eq. (\ref{MGW eff dL eq})) while not affecting the angular diameter distances and hence the time delay scaling (Eq. (\ref{MGW time delay scaling})) and time delay measurement.

\begin{figure}
\centering
\includegraphics[scale=0.47]{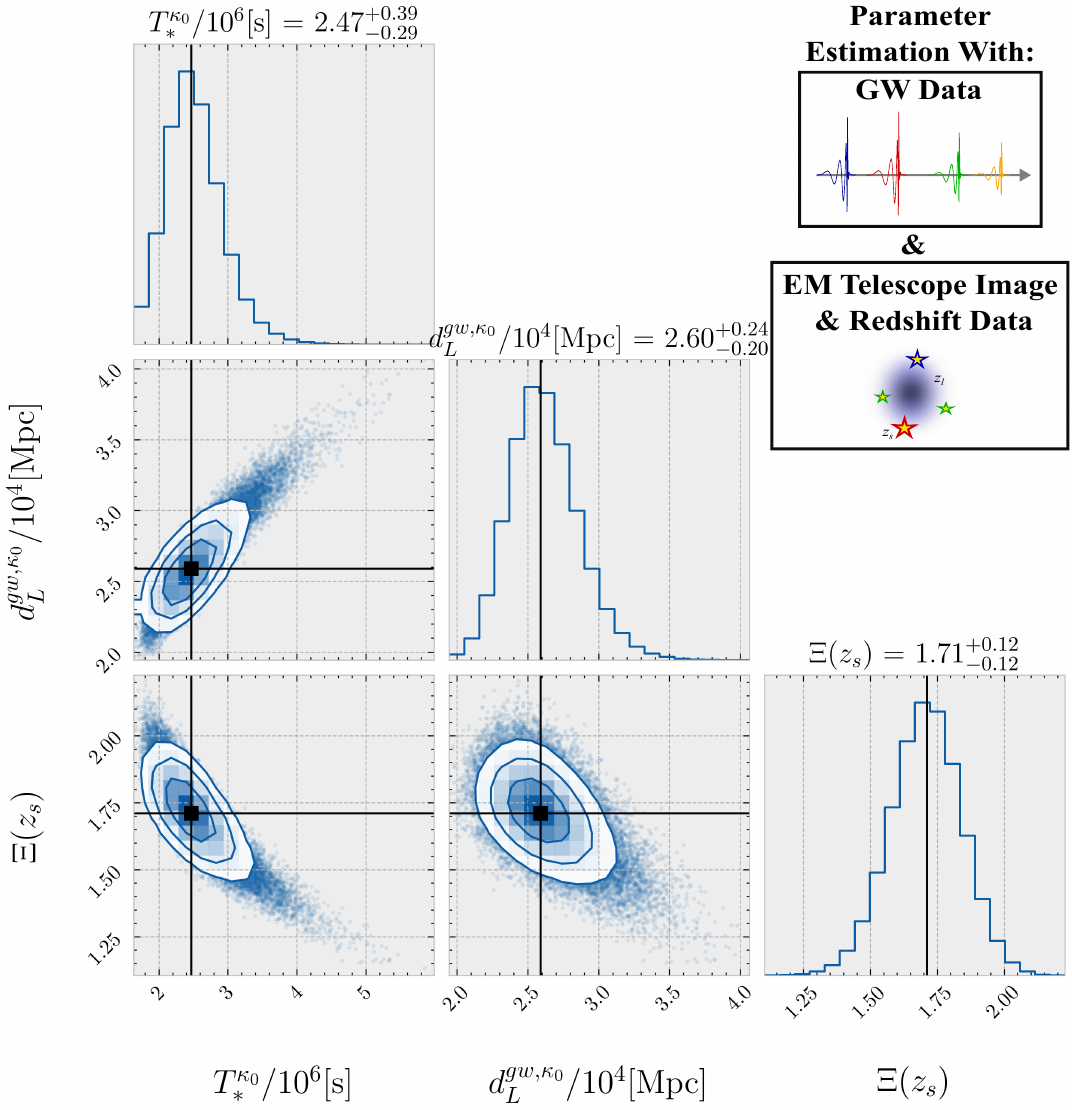}
\caption{Result of $\Xi(z_s)$ inference from lensed GW under injected parameters shown in Table \ref{table: dark siren MST lens reconstruction injection setup} (with the addition of $n = 2, \Xi_0=1.8$). After a posterior distribution of $T_*^{\kappa_0}$ and $d_L^{\text{gw}, \kappa_0}$ is obtained in the lens reconstruction run shown in Fig. \ref{MG standard PE} in the Appendix, assuming $z_s, z_l$ and $\theta_E$ are known from EM observation and lens system identification, Eq. (\ref{Xi(z_s) calculation}) is used to infer $\Xi(z_s)$. The peak of the posterior overlaps with the injection value of $\Xi(z_s) = 1.71$.}
\label{MG T vs dL vs Xi}
\centering
\end{figure}

Using the injection parameters given in Table \ref{table: dark siren MST lens reconstruction injection setup}, also injecting $n = 2, \Xi_0=1.8$, which leads to $\Xi(z_s) = 1.71$, we show the recovery of $\Xi(z_s)$ for an SIE model under unknown mass sheet transformation. The full lens reconstruction result of the free parameters $\{\vec{y}=(y_1, y_2), f, T_*^{\kappa_0}, d_L^{\text{gw}, \kappa_0}\}$ is shown in Fig. \ref{MG standard PE} in the Appendix. Assuming $z_s, z_l$ and $\theta_E$ are known from EM observation, we can infer the value of $\Xi(z_s)$ using Eq. (\ref{Xi(z_s) calculation}). The recovered posterior distribution of $T_*^{\kappa_0}$ and $d_L^{\text{gw}, \kappa_0}$ is shown in Fig. \ref{MG T vs dL vs Xi}.

\subsection{Implications on Selected Other Applications of Strongly Lensed Gravitational Waves}
As demonstrated, the modified GW propagation test is affected by the similarity transformation degeneracy but not the mass-sheet degeneracy. In short, the two degeneracies affect different strong lensing applications differently. Studies that do not require lens reconstruction should not be affected by both degeneracies. Two of such examples are testing speed of GW against light \citep{Collett_2017, Fan_2017} and associating fast radio burst sources with GW sources \citep{singh2023dejavu}: they are model independent and only require comparing the time delay between GWs and EM counterpart. In addition, GW polarization tests are also unaffected, as they rely on comparing the polarization content of the lensed GW signals that are each detected by a different detector configuration \citep{Smith_2017, Goyal_2021, hernandez2022measuring} and do not require lens modelling.

Meanwhile, cosmography studies with strongly lensed GWs, such as Hubble constant measurement as demonstrated above are affected by the similarity transformation degeneracy, as they generally require lens reconstruction to single out the effect induced by lensing and a further decoupling of the astronomical quantities (such as redshifts and the angular diameter distances) inside the time delay scaling $T_*$ and luminosity distance $d_L$. Specific applications such as Hubble constant measurement would suffer from mass-sheet degeneracy as well. Non-lensing related means from EM telescope observation, for instance, independent estimates of the lens mass using spectroscopic measurement on velocity dispersion \citep{Schneider_2013}, are required to break such degeneracy. However, \citet{Cremonese_2021}; Chen et al. (in prep.) showed that if GW lensing is in the wave optics regime, it would be possible to break the mass-sheet degeneracy from GW observation alone.

\section{Summary and Outlook}
\begin{table*}
\centering
\begin{tabular}{|c|c|c|c|c|c|c|c|c|}
 \hline
 \multicolumn{1}{|c|}{} &
 \multicolumn{5}{|c|}{Can we measure ...?} &
 \multicolumn{2}{|c|}{Can we break ... degeneracy?} &
 \multicolumn{1}{|c|}{Can we test ...?}
 \\
 
 \hline
  Observations & $\vec{y}$ & $f$ & $T_*$ & $d_L$ & $\theta_E$ & Similarity transformation & Mass-sheet  & Modified GW propagation\\
 \hline
 2 GW images & \xmark & \xmark & \xmark & \xmark & \xmark & \xmark & \xmark & \xmark \\
 \hline
 3 GW images (correct ordering) & \cmark * & \cmark * & \cmark * & \cmark * & \xmark & \xmark & \xmark & \xmark \\
 \hline
 4 GW images & \cmark & \cmark & \cmark & \cmark & \xmark & \xmark & \xmark & \xmark \\
 \hline
 4 GW images \& EM obs (w/o $z_l, z_s$) & \cmark & \cmark & \cmark & \cmark & \cmark & \xmark & \xmark & \xmark \\
 \hline
 4 GW images \& EM obs (w/ $z_l$, w/o $z_s$) & \cmark & \cmark & \cmark & \cmark & \cmark & \cmark & \xmark & \xmark \\
 \hline
 4 GW images \& EM obs (w/ $z_l, z_s$) & \cmark & \cmark & \cmark & \cmark & \cmark & \cmark & \xmark & \cmark \\
 \hline
 4 GW images \& EM obs (w/ $z_l, z_s$ \& $v$) & \cmark & \cmark & \cmark & \cmark & \cmark & \cmark & \cmark & \cmark \\
 \hline
\end{tabular}
\caption{Summary table of the measurability of SIE lens system parameters (the dimensionless observables position $\vec{y}$, the axis ratio $f$, the time delay scaling $T_*$, the luminosity distance $d_L$ and the angular Einstein radius $\theta_E$), can the similarity transformation degeneracy and the mass-sheet ($\kappa_0-H_0$) degeneracy be broken, and can the modified GW propagation test be done under seven different observation scenarios, which from top to bottom are: 1. when only two of the four lensed GWs are observed; 2. when only three of the four lensed GWs are observed, with the ordering of the three images identified or assumed correctly within the four; 3. all four lensed GWs are observed without complementary information from EM telescope (dark siren); 4. all four lensed GWs are observed and the lens system is identified in EM observation, but redshifts of lens and source are both unavailable; 5. all four lensed GWs are observed and the lens system is identified in EM observation, only redshift of lens is available; 6. all four lensed GWs are observed and the lens system is identified in EM observation, with redshifts of both lens and source available; 7. all four lensed GWs are observed and the lens system is identified in EM observation, with redshifts of both lens and source available, and the lens galaxy's velocity dispersion can be measured.\\
\cmark: can measure/break the degeneracy/test.\\
*: can measure, but with large uncertainty.\\
\xmark: cannot measure/break the degeneracy/test.
}
\label{table: summary table}
\end{table*}

\begin{table*}
\centering
\begin{tabular}{|c|c|c|}
 \hline
 \multicolumn{1}{|c|}{} &
 \multicolumn{2}{|c|}{Can we break ... degeneracy?}
 \\
 
 \hline
  Observations (all four images observed) & Similarity transformation & Mass-sheet\\
 \hline
 Dark standard siren (GW) & \xmark & \xmark \\
 \hline
 Bright standard siren (GW \& EM obs w/ $z_l, z_s$, w/o $v$) & \cmark & \xmark \\
 \hline
 Bright standard siren (GW \& EM obs w/ $z_l, z_s, v$) & \cmark & \cmark \\
 \hline
 Standard Candle (w/ $z_l, z_s$, w/o $v$) & \cmark & \xmark \\
 \hline
 Standard Candle (w/ $z_l, z_s, v$) & \cmark & \cmark \\
 \hline
 Time-varying, non-standard candle light source (w/ $z_l, z_s$, w/o $v$) & \cmark & \xmark \\
 \hline
 Time-varying, non-standard candle light source (w/ $z_l, z_s, v$) & \cmark & \cmark \\
 \hline
 Non-time-varying, non-standard candle light source (w/ $z_l, z_s, v$) & \cmark * & \xmark \\
 \hline
\end{tabular}
\caption{Summary table of can the similarity transformation degeneracy and the mass-sheet degeneracy be broken under eight different lensing of GW and/or lensing of light observation scenarios (which we assumed a SIE lens system, with all four images are observed for all cases). From top to bottom are: 1. dark siren lensed GWs are observed; 2. lensed GWs observation with complementary EM information (redshift of source and lens) (\textit{i.e.} bright siren); 3. lensed GWs are observed with complementary EM information (redshift of source and lens), with the addition of lens galaxy velocity dispersion $v$ measurement; 4. lensed standard candle (such as supernovae) are observed, with redshift measurement of source and lens; 5. lensed standard candle (such as supernovae) are observed, with measurement of redshift of source and lens, as well as velocity dispersion $v$ of lens galaxy; 6. lensing of a time-varying, but non-standard candle light source is observed, with redshift measurement of source and lens; 7. lensing of a time-varying, but non-standard candle light source is observed, with measurement of redshift of source and lens, as well as velocity dispersion $v$ of lens galaxy; 8. lensing of a non-time-varying and non-standard candle light source is observed, with measurement of redshift of source and lens, as well as velocity dispersion $v$ of lens galaxy.\\
\cmark: can break the degeneracy.\\
*: can break the degeneracy, but $H_0$ cannot be measured even neglecting mass sheet.\\
\xmark: cannot measure/break the degeneracy.
}
\label{table: summary table 2}
\end{table*}

We have presented a framework for performing lens reconstruction based on dark siren strongly lensed GWs using a Bayesian analysis framework and "irreducible" lensing parameters. GW-only lensing will always suffer from two major degeneracies, which we have detailed in this work. The first one is the similarity transformation degeneracy, which is a degeneracy between the lens and source redshift, as well as Einstein radius of the lens and cosmological parameters such as the Hubble constant. The second one is the mass-sheet degeneracy, which can alter the lens model and together with modification on the Hubble constant, they can keep the lensing observables unchanged. Not accounting for the mass-sheet degeneracy will lead to biased $H_0$ measurements. 

The similarity transformation degeneracy can be broken once we have a EM counterpart of the GW signal or we localize and identify the corresponding lens system (when there is no EM counterpart from the GW event) as demonstrated in \citet{Hannuksela_2020}. Unfortunately, the mass-sheet degeneracy will not be broken in our strong lensing case even if we have the redshift information and the Einstein radius from high-resolution telescope images. Instead, other observation such as spectroscopic measurement on lens galaxy's velocity dispersion is needed to break the degeneracy \citep{Schneider_2013}.

Furthermore, we have also reviewed two GW lensing science applications and how they are affected by these degeneracies. Table \ref{table: summary table} summarizes the quantities that we can measure (in an SIE lens system), the degeneracies that we can break, and if modified GW dispersion relations can be tested for different observation scenarios, such as dark siren GW observation and GW observation complemented by redshift and lens galaxy's velocity dispersion measurement. In Table \ref{table: summary table 2}, we also summarized whether similarity transformation and mass-sheet degeneracy can be broken under eight different observational scenario of GW and/or EM lensing. In addition, for Hubble constant measurements, one advantage of GW strong lensing is that one can take advantage of the millisecond accuracy of arrival time measurements afforded by the detectors \citet{Fan_2017}, which can complement the more traditional time-delay measurements \citep{refId0}. 
For this Hubble constant measurement, one needs to have knowledge of the lens redshift and size/Einstein radius, but the source redshift is, intriguingly, not required.

In addition, we also summarised the resolvability of the similarity transformation and mass-sheet degeneracy under different scenarios of observation of GW and/or light sources (such as standard candle, non-standard candle but time-varying light sources) in Table \ref{table: summary table 2.}.

Secondly, we also showed the modified GW propagation test is immune to mass-sheet degeneracy (and does not require knowledge on Hubble constant), which can be beneficial for testing theories modifying general relativity. 

Besides, understanding these GW-based lens reconstructions and their degeneracies will be important also for other science applications, for instance, broader cosmography study. Understanding the galaxy modelling capabilities will also play a role in understanding how GWs will help probe dark matter and substructure within galaxy lenses. Indeed, we hope that our work will help motivate further studies into complementary avenues in multimessenger lensing, where GWs and EM can provide complementary information that will help in understanding of strong lenses and forming a rich scientific case.

\section{Acknowledgements}

We thank Hemantakumar Phurailatpam, Laura Uronen, Leo Ng for valuable discussions, feedback and comments. 
Jason S.C. Poon is supported by the Hong Kong PhD Fellowship Scheme (HKPFS) from the Hong Kong Research Grants Council (RGC). 
Jason S.C. Poon and Otto A. Hannuksela acknowledge support by grants from the ResearchGrants Council of Hong Kong (Project No. CUHK 14304622 and 14307923), the start-up grant from the Chinese University of Hong Kong, and the Direct Grant for Research from the Research Committee of The Chinese University of Hong Kong.
Stefano Rinaldi acknowledges financial support from the European Research Council for the ERC Consolidator grant DEMOBLACK, under contract no. 770017 and from the EU Horizon 2020 Research and Innovation Program under the Marie Sklodowska-Curie Grant Agreement No. 734303. Justin Janquart and Harsh Narola are supported by the research programme of the Netherlands Organisation for Scientific Research (NWO). 

\section{Data Availability}
The data underlying this article will be shared on reasonable request
to the corresponding authors.

\bibliographystyle{mnras}
\bibliography{main}

\appendix
\section{More Detailed Plots for Parameter Estimation Results Presented}

\begin{figure*}
\centering
\includegraphics[scale=0.6]{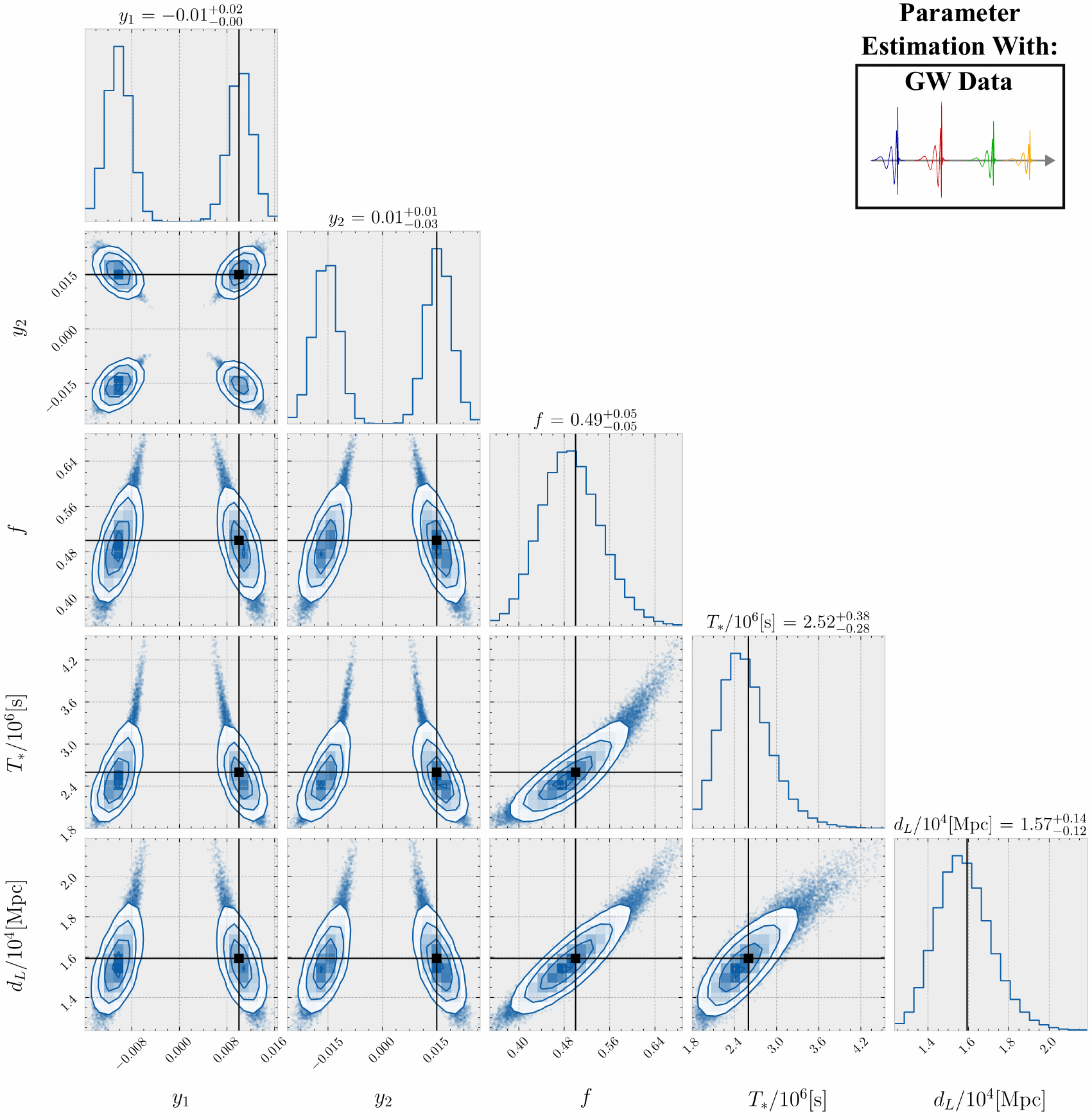}
\caption{Result of SIE lens reconstruction of dark siren lensed GW using injected parameters shown in Table \ref{table:dark_siren_lens_reconstruction_injection_setup} and Table \ref{table:dark_siren_lens_reconstruction_injected_observables}, which a Gaussian likelihood with $5\%$ error on observed relative time delays and effective luminosity distances is assumed. After taking similarity transformation degeneracy into account, the five free parameters are $\{\vec{y}=(y_1, y_2), f, T_* \text{ (in seconds)}, d_L \text{ (in Mpc)}\}$, which are the dimensionless source position, the SIE axis ratio, the time delay scaling $T_*\equiv\frac{1+z_l}{c}\frac{D_l D_s}{D_{ls}} \theta_E^2$ and the (unlensed) luminosity distance respectively. We can see, all parameters are recovered fairly well and except from the dimensionless source position which suffers from multiple solutions (as SIE lens has two degrees of reflectional symmetry), there is no degeneracy among the parameters.
}
\label{fig:STD_standard_PE}
\centering
\end{figure*}

\begin{figure*}
\centering
\includegraphics[scale=0.44]{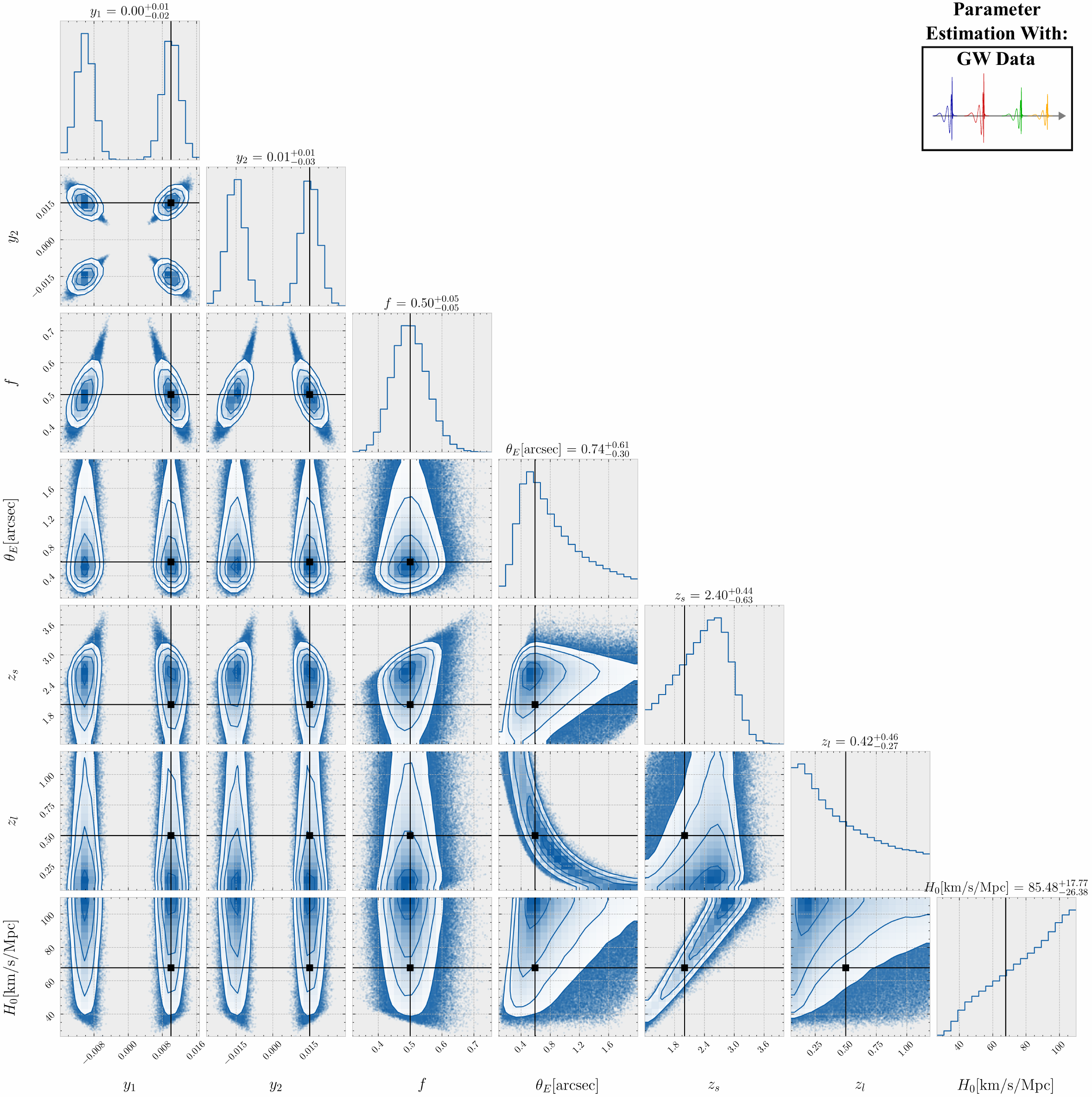}
\caption{Result of SIE lens reconstruction of dark siren lensed GW using injected parameters shown in Table \ref{table:dark_siren_lens_reconstruction_injection_setup} and Table \ref{table:dark_siren_lens_reconstruction_injected_observables}, without accounting for similarity transformation degeneracy. A Gaussian likelihood with $5\%$ error on observed relative time delays and effective luminosity distances is assumed. The seven free parameters are $\{\vec{y}=(y_1, y_2), f, \theta_E \text{ (in arcseconds)}, z_s, z_l, H_0 \text{ (in km} \text{ s}^{-1} \text{ Mpc}^{-1}\text{)}\}$, which are the dimensionless source position, the SIE axis ratio, the Einstein radius, the source and lens redshift and the Hubble constant respectively. While the dimensionless part of the system (\textit{i.e.} $\vec{y}$ and $f$) can be recovered successfully,  all other parameters that describe the absolute scale of the system (\textit{i.e.} $\theta_E, z_s, z_l$ and $H_0$) are poorly recovered (which can be sensitive to prior choice) and showed high degree of degeneracies, which is expected from similarity transformation degeneracy.}
\label{fig:STD_degenerate_case_full_PE}
\centering
\end{figure*}

\begin{figure*}
\centering
\includegraphics[scale=0.6]{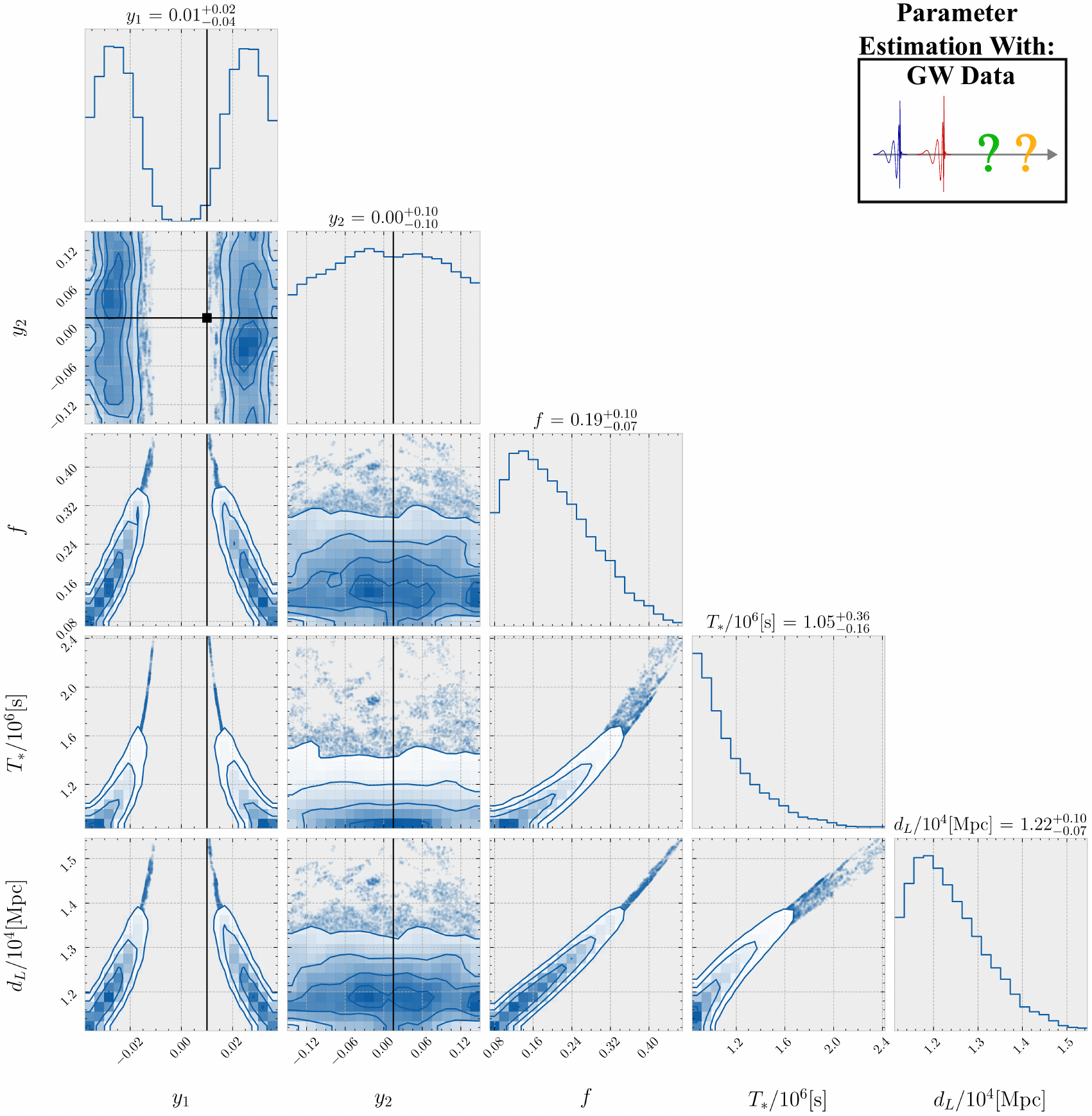}
\caption{Result of SIE lens reconstruction of dark siren lensed GW using injected parameters shown in Table \ref{table:dark_siren_lens_reconstruction_injection_setup} and the observables of the first two images from Table \ref{table:dark_siren_lens_reconstruction_injected_observables}, which a Gaussian likelihood with $0.001\%$ error on observed relative time delays and effective luminosity distances is assumed. After taking similarity transformation degeneracy into account, the five free parameters are $\{\vec{y}=(y_1, y_2), f, T_* \text{ (in seconds)}, d_L \text{ (in Mpc)}\}$, which are the dimensionless source position, the SIE axis ratio, the time delay scaling $T_*\equiv\frac{1+z_l}{c}\frac{D_l D_s}{D_{ls}} \theta_E^2$ and the (unlensed) luminosity distance respectively. We can see, the parameters cannot be recovered and the posterior is uninformative.
}
\label{STD 2 img standard PE}
\centering
\end{figure*}

\begin{figure*}
\centering
\includegraphics[scale=0.6]{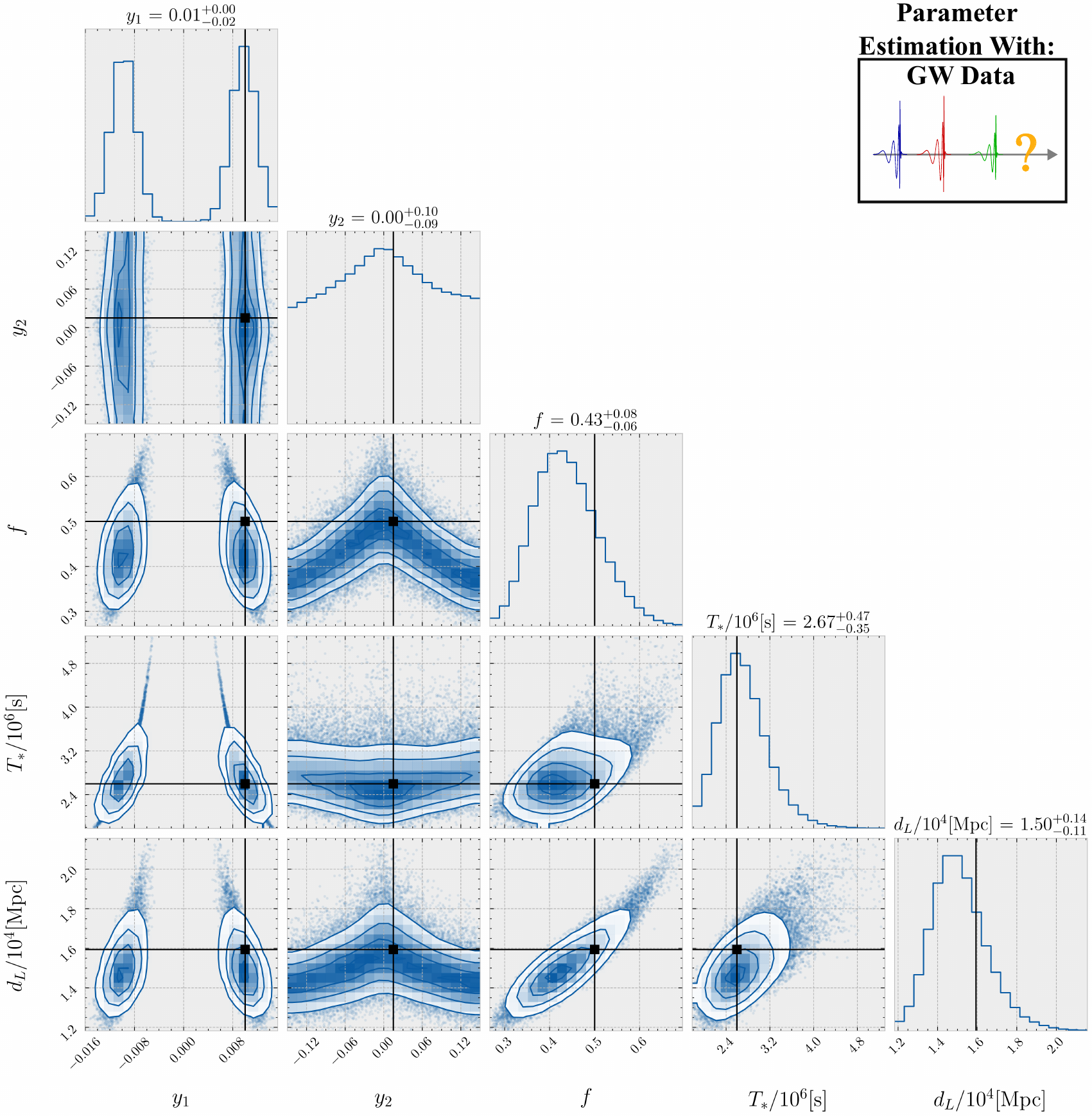}
\caption{Result of SIE lens reconstruction of dark siren lensed GW using injected parameters shown in Table \ref{table:dark_siren_lens_reconstruction_injection_setup} and the observables of the first three images from Table \ref{table:dark_siren_lens_reconstruction_injected_observables}, which a Gaussian likelihood with $5\%$ error on observed relative time delays and effective luminosity distances is assumed. After taking similarity transformation degeneracy into account, the five free parameters are $\{\vec{y}=(y_1, y_2), f, T_* \text{ (in seconds)}, d_L \text{ (in Mpc)}\}$, which are the dimensionless source position, the SIE axis ratio, the time delay scaling $T_*\equiv\frac{1+z_l}{c}\frac{D_l D_s}{D_{ls}} \theta_E^2$ and the (unlensed) luminosity distance respectively. We can see, the parameters can be recovered, but with very large uncertainty.
}
\label{STD 3 img standard PE (5percent error)}
\centering
\end{figure*}

\begin{figure*}
\centering
\includegraphics[scale=0.6]{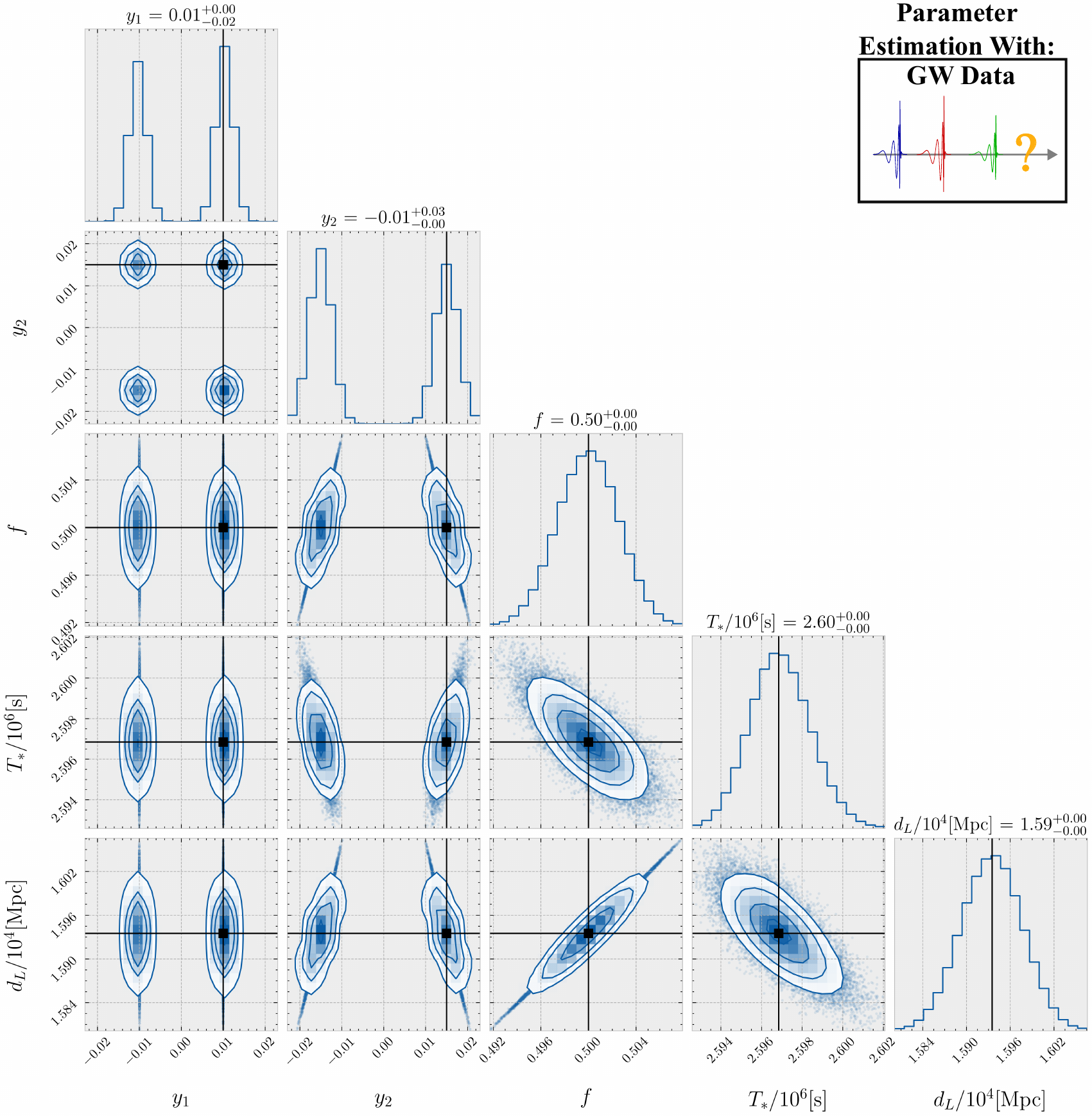}
\caption{Result of SIE lens reconstruction of dark siren lensed GW using injected parameters shown in Table \ref{table:dark_siren_lens_reconstruction_injection_setup} and the observables of the first three images from Table \ref{table:dark_siren_lens_reconstruction_injected_observables}, which a Gaussian likelihood with $0.01\%$ error on observed relative time delays and effective luminosity distances is assumed. After taking similarity transformation degeneracy into account, the five free parameters are $\{\vec{y}=(y_1, y_2), f, T_* \text{ (in seconds)}, d_L \text{ (in Mpc)}\}$, which are the dimensionless source position, the SIE axis ratio, the time delay scaling $T_*\equiv\frac{1+z_l}{c}\frac{D_l D_s}{D_{ls}} \theta_E^2$ and the (unlensed) luminosity distance respectively. We can see, all parameters are recovered fairly well and except from the dimensionless source position which suffers from multiple solutions (as SIE lens has two degrees of reflectional symmetry), there is no degeneracy among the parameters.
}
\label{STD 3 img standard PE (0.01percent error)}
\centering
\end{figure*}

\begin{figure*}
\centering
\includegraphics[scale=0.6]{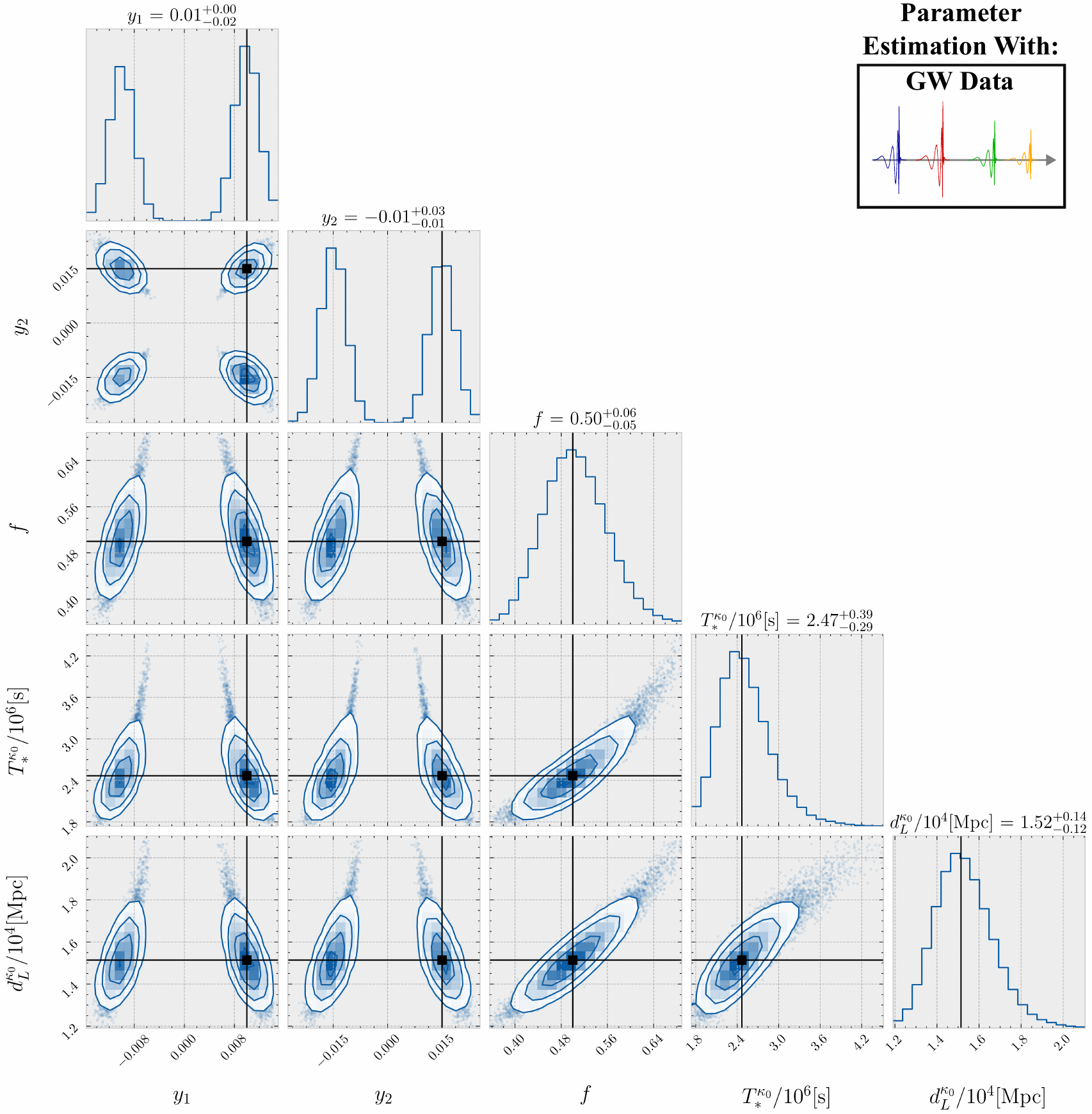}
\caption{Result of SIE lens reconstruction of dark siren lensed GW using injected parameters shown in Table \ref{table: dark siren MST lens reconstruction injection setup} and Table \ref{table: dark siren MST lens reconstruction injected observables}, which a Gaussian likelihood with $5\%$ error on observed relative time delays and effective luminosity distances is assumed. After taking similarity transformation degeneracy and mass-sheet degeneracy into account, the five free parameters are $\{\vec{y}=(y_1, y_2), f, T_*^{\kappa_0} \text{ (in seconds)}, d_L^{\kappa_0} \text{ (in Mpc)}\}$, which are the dimensionless source position (before mass sheet transformation), the SIE axis ratio, the mass-sheet-degenerated time delay scaling $T_*^{\kappa_0}\equiv (1-\kappa_0) \frac{1+z_l}{c}\frac{D_l D_s}{D_{ls}} \theta_E^2$ and the mass-sheet-degenerated luminosity distance respectively. We can see, all parameters are recovered fairly well and except from the dimensionless source position which suffers from multiple solutions (as SIE lens has two degrees of reflectional symmetry), there is no degeneracy among the parameters.}
\label{MSD standard PE}
\centering
\end{figure*}

\begin{figure*}
\centering
\includegraphics[scale=0.6]{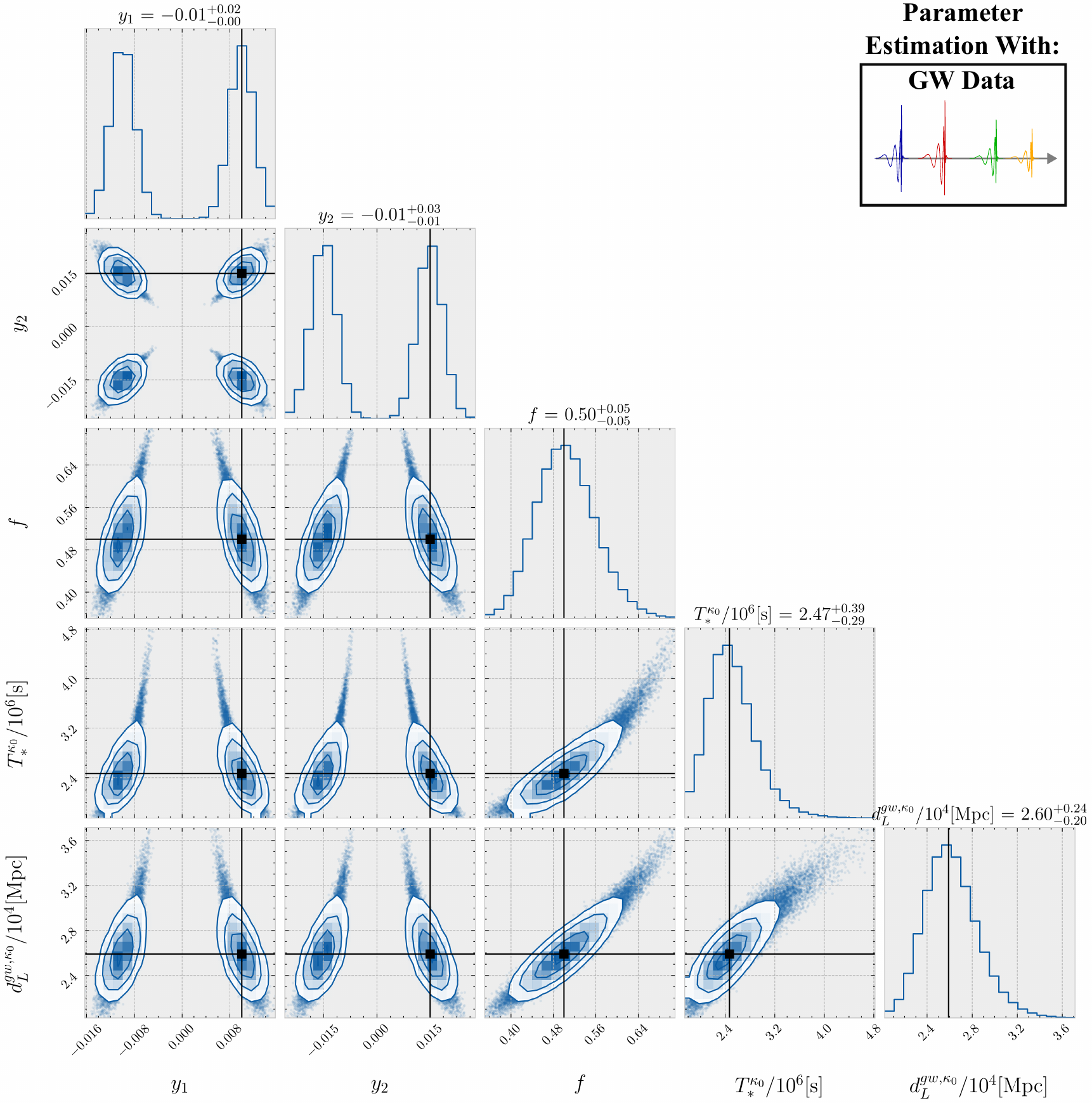}
\caption{Result of SIE lens reconstruction of dark siren lensed GW using injected parameters shown in Table \ref{table: dark siren MST lens reconstruction injection setup}, with the addition of modified GW propagation effect of $n=2$ and $\Xi_0=1.8$. A Gaussian likelihood with $5\%$ error on observed relative time delays and effective luminosity distances is assumed. After taking similarity transformation degeneracy, mass-sheet degeneracy and modified GW propagation into account, the five free parameters are $\{\vec{y}=(y_1, y_2), f, T_*^{\kappa_0} \text{ (in seconds)}, d_L^{\text{gw}, \kappa_0} \text{ (in Mpc)}\}$, which are the dimensionless source position (before mass sheet transformation), the SIE axis ratio, the mass-sheet-degenerated time delay scaling $T_*^{\kappa_0}\equiv (1-\kappa_0) \frac{1+z_l}{c}\frac{D_l D_s}{D_{ls}} \theta_E^2$ and the mass-sheet-degenerated GW luminosity distance respectively. All parameters are recovered well and there is no degeneracy.}
\label{MG standard PE}
\centering
\end{figure*}

\bsp	
\label{lastpage}
\end{document}